%% file: main.tex
\documentclass[10pt, conference, compsocconf,letterpaper]{IEEEtran}
\ifCLASSINFOpdf
\else
\fi
\usepackage[cmex10]{amsmath}

\usepackage{mdwlist}

\usepackage{comment, graphicx, subfig, multirow, color}
\usepackage{algorithm2e}
\usepackage{url, cite}
\begin{document}

\title{Octopus: A Secure and Anonymous DHT Lookup}


%\begin{comment}
\author{\IEEEauthorblockN{Qiyan Wang}
\IEEEauthorblockA{Department of Computer Science\\
University of Illinois at Urbana-Champaign\\
IL, U.S.A\\
qwang26@illinois.edu}
\and
\IEEEauthorblockN{Nikita Borisov}
\IEEEauthorblockA{Department of Electrical and Computer Engineering\\
University of Illinois at Urbana-Champaign\\
IL, U.S.A.\\
nikita@illinois.edu}
}
%\end{comment}


%Telephone: (800) 555--1212, Fax: (888) 555--1212}


\maketitle

\begin{abstract}

Distributed  Hash Table (DHT) lookup is a core technique in structured peer-to-peer (P2P) networks.  Its decentralized nature introduces security and
privacy vulnerabilities for applications built on top of them; we thus set out to design a lookup mechanism achieving both security and anonymity, heretofore an open problem.  We present Octopus, a novel DHT lookup which provides strong guarantees for both security and anonymity.  Octopus uses attacker identification mechanisms to discover and remove malicious nodes, severely limiting an adversary's ability to carry out active attacks, and splits lookup queries over separate anonymous paths and introduces dummy queries to achieve high levels of anonymity.  We analyze the security of Octopus by developing an event-based simulator to show that the attacker discovery mechanisms can rapidly identify malicious nodes with low error rate. We calculate the anonymity of Octopus using probabilistic modeling and show that Octopus can achieve near-optimal anonymity. We evaluate Octopus's efficiency on Planetlab with 207 nodes and show that Octopus has reasonable lookup latency and manageable communication overhead.

\end{abstract}


% \fi
\IEEEpeerreviewmaketitle

\input{introduction.tex}

\input{sys_model.tex}

\input{secure_lookup.tex}

\input{anonymous_lookup.tex}

\input{experiment.tex}

\input{background.tex}	% background & related work

\input{conclusion.tex}

\section{Acknowledgements}

We would like to thank Prateek Mittal for helpful discussion throughout the project. We also thank George Danezis for invaluable feedback on the earlier draft of the paper. In addition, we are grateful to Prateek Mittal for sharing his codes of the event-based simulator and the DHT implementation on Planetlab. Finally, we thank anonymous reviewers for helpful comments to improve the paper. This research was supported in part by NSF CNS 09-53655.


%\section*{Acknowledgment}

%The authors would \cite{IEEEexample:articleetal} like to thank...
%more thanks here



%\bibliographystyle{IEEEtran}
%\begin{thebibliography}{1}
%\end{thebibliography}

\bibliographystyle{IEEEtran}
\bibliography{IEEEabrv,sigproc}

\input{appendix.tex}

\end{document}

%% file: introduction.tex
\section{Introduction}
\label{sec:introduction}

Structured peer-to-peer networks, such as Chord~\cite{chord} or Kademlia~\cite{Kad}, allow the creation of scalable distributed applications that can support millions of users.  They have been used to build a number of successful applications, including P2P file sharing (Overnet, Kad, Vuze DHT\footnote{\url{http://www.vuze.com/}}) and content distribution (CoralCDN~\cite{freedman+:nsdi04}), and many others have been proposed, such as distributed file systems~\cite{rowstron-druschel:sosp01}, anonymous communication systems~\cite{AP3,salsa,NISAN,torsk,shadowwalker}, and online social networks~\cite{shakimov+:wosn09,social_cutillo}.  At the heart of these networks lies a distributed hash table (DHT) lookup mechanism that implements a decentralized key--value store.  The DHT allows efficient storage and coordination among very large collections of nodes; however, its decentralized nature creates a number of security and privacy vulnerabilities.  Because peers have to rely on other peers to determine the state of the network, malicious nodes could provide misinformation to misdirect an honest user's lookups~\cite{wallach:ssts03}.  Likewise, nodes can profile the lookup activities of other nodes and learn what files or websites they are interested in or who their friends may be. Recent research shows that anonymous communication systems based on anonymity-deficient DHT lookups have severe vulnerabilities to information leak attacks~\cite{prateek08, qiyan10}.

To address these issues, our goal is to design a lookup mechanism that achieves both security and anonymity (where anonymity means no information is revealed about which nodes are looking up which values), heretofore an open problem.  We note that security is a {\it necessary condition} for anonymity, because without security, malicious nodes can misdirect the lookup towards colluding nodes and learn about the lookup target. 
On the other hand, security is {\it not a sufficient condition} for anonymity. Some existing lookup schemes designed to resist active attacks are not suitable to build anonymous DHT systems, e.g., due to heavily relying on redundant transmission that leaks information about the lookup initiator and/or target~\cite{castro, Halo, RCP}. Furthermore, even with a secure lookup that itself does not cause information leak, an inappropriate design of the DHT system can still lead to anonymity vulnerabilities~\cite{qiyan10}.

Our contributions in this work include:

1) We propose a suite of novel security mechanisms--{\it attacker discovery} for DHT systems. Our mechanisms proactively identify and remove malicious peers. We develop an event-based simulator to show that our identification mechanism is capable of rapidly discovering malicious nodes with low error rate. For a network with 20\% malicious nodes, it can correctly identify all attacking nodes within 30 minutes. We compare our scheme with Halo~\cite{Halo}, a state-of-the-art secure DHT system, and show that our scheme provides better robustness against active attacks.
Furthermore, our design does not rely on redundant transmission, and thus is suitable to construct anonymous DHT systems.

2) With the proposed security mechanisms, we put forward a secure and anonymous DHT lookup {\it Octopus}. Octopus splits individual queries used in a lookup over multiple anonymous paths, and introduces dummy queries, to make it difficult for an adversary to learn the eventual target of a lookup.  Unlike most previous works that only analytically evaluate the systems' anonymity, we use probabilistic modelling with the help of simulation to calculate the information leak, so that users can know how much anonymity can be actually provided by the system. We show that Octopus provides near-optimal anonymity for both the lookup initiator and target. In a network of 100\,000 nodes with 20\% malicious nodes, Octopus only leaks 0.57 bit of information about the initiator and 0.82 bit of information about the target; these are 6 times and 4 times better than what previous works~\cite{NISAN, torsk} were able to achieve, respectively. 

3) For performance evaluation, we measure the lookup latency of Octopus on Planetlab with 207 nodes, and compare it with the base-line scheme Chord~\cite{chord} and Halo~\cite{Halo}. The lookup latency of Octopus is comparable to that of Chord, and even better than that of Halo. While Octopus incurs relatively higher communication overhead than Chord and Halo to provide extra security and/or anonymity guarantees, the bandwidth consumption of Octopus is still manageable, which is only a few kbps for each node.

The remainder paper is organized as follows. Section~\ref{sec:model} presents the system model. We describe our security and anonymity mechanisms in Section~\ref{sec:secure_lookup} and Section~\ref{sec:anony_lookup}. The efficiency evaluation is provided in Section~\ref{sec:performance} and Section~\ref{sec:related_work} presents the related work. We conclude in Section \ref{sec:conclusion}.

%% file: sys_model.tex
\section{System Model}
\label{sec:model}

\subsection{Threat Model}

In the same vein of related works~\cite{AP3, salsa, NISAN, torsk, shadowwalker, RCP, RCP_pq}, we do not consider a global adversary that is capable of controlling the whole network and observing all communication traffic. Such a global adversary seems unpractical in large-scaled P2P networks. Instead, we assume a partial adversary that controls a fraction $f$ of all nodes in the network ($f$ is typically assumed to be up to 20\%).  Malicious nodes can behave in an arbitrarily malicious way, such as intercepting, modifying or dropping any messages going through them, or injecting fake messages to any other nodes . We also assume that malicious nodes can log any messages they have seen and access to a high-speed communication channel to share any information with very low transmission delay.

Also similar to related works, we do not attempt to solve the problem of Sybil attack~\cite{sybil} in this work. Defending sybil attack is an interesting research area that has drawn a lot of attentions; a number of effective solutions have been proposed, such as~\cite{nikita_sybil, Sybil_Geroge_09, Sybillimit}. All these solutions are applicable to Octopus as extensions to resist Sybil attacks. 

\subsection{Design Goals}

The major goal for security is to avoid lookups being biased by malicious nodes. In other words, given a lookup target, it should not be possible to misdirect the lookup path or bias the final lookup result.

Pfitzmann and Hansen defined several relevant anonymity properties for message-based communication, such as sender and receiver anonymity~\cite{anon_terminology}.  We consider equivalent properties in the context of DHT lookups.
\begin{itemize}
	\item \emph{Initiator anonymity}: given a lookup target, it should not be possible to determine its initiator. 	
	\item \emph{Target anonymity}: given a lookup initiator, it should not be possible to determine its target. 	
	\item \emph{Query unlinkability}: given several queries with known targets, it should not be possible to find out if they came from the same initiator.
\end{itemize}

%% file: secure_lookup.tex
\section{Security Mechanisms of Octopus}
\label{sec:secure_lookup}

\subsection{Problem Description}

In a DHT system, like Chord~\cite{chord}, each node is assigned a unique ID associated with its IP address, and owns the IDs from itself to its direct predecessor on the ring. Each node $X$ maintains a list of $\Theta(\log N)$ contact nodes (called {\it fingers}), where $N$ is the network size, and the $i$-th finger of $X$ is the owner (or successor) of the ID $id_X + 2^{i-1}$ (the first finger, i.e., $i=1$, is $X$'s direct successor). Besides, each node maintains a list of successor nodes for stabilization. Some DHTs~\cite{Pastry} also utilize the list of successors during lookups to speed up the lookup process in the last few hops. 

We study the more general case where the successor lists are used in lookups. We refer to the combination of the fingertable and the successor list as the  {\it routing table}. More specifically, we consider the following lookup procedure. Assuming a lookup initiator $I$ wants to find the owner of value $v$, it first queries $n_1$, the node that is closest to $v$ in its routing table, and $n_1$ sends its routing table to $I$ \footnote{In vanilla DHT lookups, $I$ tells $v$ to each queried node, which will return the finger closest to $v$; however, this reveals the lookup target to malicious intermediate nodes. Hence, for anonymous lookups, $I$ asks each intermediate node for its full routing table, without revealing $v$.}. Then out of $n_1$'s routing table, $I$ finds the node $n_2$ closest to $v$ and asks $n_2$ for its routing table. The process iteratively proceeds until reaching $n_k$, for which one of its successors is the owner of $v$ (i.e., the lookup target). 

Some of the queried nodes could be malicious and attempt to launch the following attacks.

\subsubsection{Lookup Bias Attack}
If the last queried node $n_k$ is malicious, it can replace the honest nodes in its successor list with malicious nodes, so that one of its (malicious) ``successors'' will be concerned as the lookup target.

\subsubsection{Lookup Misdirection Attack}
Instead of trying to bias the lookup result, malicious nodes could attempt to make $I$ query more malicious nodes during the lookup, by providing manipulated fingertables (i.e., replacing honest fingers with malicious nodes). This attack is a big threat to anonymity since the adversary can learn more information about the lookup target from a larger number of queried malicious nodes~\cite{qiyan10}.

\subsubsection{Finger Pollution Attack}
In a DHT system, each node periodically performs lookups to update its fingers. An attack related to this is that malicious nodes could attempt to pollute honest nodes' fingertables during the finger-update lookups, so that the polluted fingertables can contribute to the lookup bias and misdirection attacks.

\subsection{Security Mechanisms}

Many existing secure DHT designs~\cite{castro, Halo, salsa} employ redundant queries or lookups to tolerate misinformation provided by malicious nodes. However, the redundant transmission creates more opportunities for an adversary to gain information about the lookup initiator and/or target~\cite{prateek08}. Some schemes~\cite{RCP, RCP_pq} utilize quorum-based topologies and threshold cryptography to limit byzantine adversaries, but they also require the initiator to contact multiple nodes at each step of the lookup (for cryptographic operations), which accelerates information leak. Some other schemes, such as Myrmic~\cite{Myrmic}, prevent routing table manipulation by introducing a central trusted authority to sign each node's routing table; however, this approach is impractical since for each node join or churn, the authority has to regenerate the signatures for all related nodes, rendering a performance bottleneck.

To effectively limit active attacks while minimizing information leak, we propose a new defense strategy by letting each (honest) node secretly check the correctness of other nodes' routing tables. Such checks are preformed offline (i.e., independent of lookups), and thus do not reveal any information of lookups. To punish discovered malicious nodes, we use a certificate authority (CA) to issue certificates and revoke certificates from identified malicious nodes so that the malicious nodes can be gradually removed from the system. We note that the CA in our case is fundamentally different from that of Myrmic~\cite{Myrmic}. The latter is required to be online all the time and needs to update signatures for multiple nodes for each node churn/join. Whereas, the certificates in our scheme are independent of nodes' routing states and thus do not need to be updated frequently. Our simulation results show that the workload of our CA is sufficiently low and can be handled by most Internet servers.

There has been several fairly efficient and scalable revocation mechanisms in the literature, such as Merkle Hash Tree based certificate revocation~\cite{revocation-tree}, efficient distribution of revocation information over P2P networks~\cite{revocation-p2p}, and scalable PKI based on P2P systems~\cite{PKI-p2p}. Since certificate management in our scheme is essentially the same as these systems, we do not particularly study certificate revocation in this work. 

%%%%%%

\subsubsection{Secret Neighbor Surveillance}

To limit the lookup bias attack,  we propose secret neighbor surveillance, a mechanism that prevent malicious neighbors from manipulating their successor lists. %The key idea is to let nodes detect misbehaviors of their neighbors. 

In particular, we let each node maintain a predecessor list, in the same way as maintaining the successor list (i.e., periodically running Chord stabilization protocol anti-clockwise). The predecessor list is of the same size as the successor list, and thus each node $X$ should be contained in the successor list of any of its predecessors. In other words, if $X$ is not contained in the successor list of its predecessor, it means this predecessor is trying to manipulating its successor list by replacing $X$ with another node. Our goal is to let $X$ detect this.

\begin{figure}[h]
	\centering
	\includegraphics[height=3.7cm] {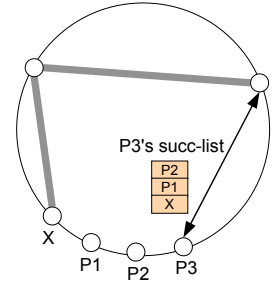}
	\caption{\small Secret neighbor surveillance. $X$ is checking if itself is included in its predecessor $P_3$'s successor list. If not, it means $P_3$ manipulates its successor list by replacing $X$ with another node.}
\label{fig:neighbor_surveillance}
\end{figure}

In the lookup bias attack, a malicious node provides a manipulated successor list in response to lookup queries. Therefore, we let $X$ {\it anonymously} sends a ``lookup query'' to one of its predecessors, say $P_3$ (see Figure~\ref{fig:neighbor_surveillance}), and checks if itself is included in $P_3$'s successor list.  ``Anonymously'' is required since if the malicious node can distinguish a testing query from real lookup queries based on the querier's identity, it can always provide the correct successor list for testing queries to avoid being detected. The anonymous transmission can be achieved by using the basic onion routing technique~\cite{onion-routing}, i.e., $X$ chooses two randomly peers as relays to forward its query to $P_3$ while using onion encryption to ensure each hop on the forwarding path can only know its previous and next hops. The two relay nodes can be found by performing a $l$-hop random walk on the overlay network (where $l=\Theta(\log N)$). The details of the random walk are provided in Appendix~\ref{ssec:random_walk}.

$X$ performs the above checks from time to time (i.e., with time interval $t_c \in_R (0, T_m]$ where $T_m$ is the maximum checking interval) on randomly selected successors. A detected malicious node will be reported to the CA. To provide a non-repudiation proof on a manipulated successor list (i.e., verifiable to the CA), we let each node sign its routing table and attach a time stamp to it.

\begin{figure}[h]
	\centering
	\includegraphics[height=3.5cm] {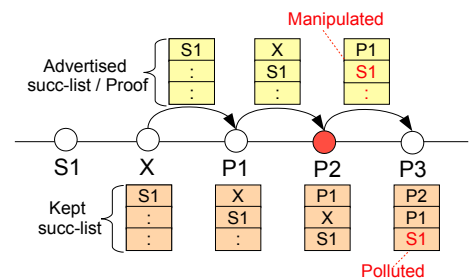}
	\caption{\small Successor list pollution. The malicious successor $P_2$ pollutes $P_3$'s successor list in $P_3$'s stabilization by providing a manipulated successor list.}
\label{fig:succlist_pollution}
\end{figure}

Another strategy to launch the lookup bias attack is to pollute honest nodes' successor lists during stabilization. For example, as shown in Figure \ref{fig:succlist_pollution}, assuming $P_2$ is malicious and $P_3$ is honest, $P_2$ can send $P_3$ a manipulated successor list excluding $X$ during $P_3$'s stabilization, so that $P_3$ will concern $X$ as dead and remove $X$ from its successor list; consequently, $P_3$ will be mistakenly identified as a malicious successor by $X$ and $P_2$ will still be uncovered. To deal with this, we let each node sign its successor list used in stabilization; also, each node keeps a queue of latest received successor lists in stabilization as {\it proof}, to prove that its successor list is not intentionally manipulated. For example, $P_3$ can provide its proof to the CA showing that its successor list is correctly computed according to the information provided by $P_2$. If $P_3$'s proof is verified (according to the stabilization algorithm) by the CA, then the suspicion on $P_3$ is cleared and the CA will request $P_2$ for its proof, and check it against $P_3$'s proof. This process is repeated until finding a node that cannot provide valid proofs and this node is then judged as the malicious node.

\subsubsection{Secret Finger Surveillance}

Likewise, we propose a {\it secret finger surveillance} mechanism for the lookup misdirection attack by limiting malicious nodes from manipulating their fingertables in the lookup. %This mechanism essentially lets each node secretly check the integrity of others' fingertables. 

\begin{figure}[h]
	\centering
	\includegraphics[height=4.0cm] {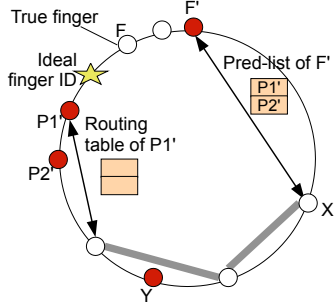}
	\caption{\small Secret finger surveillance. $X$ is checking if node $Y$ has replaced its finger $F$ with a malicious node $F'$. $X$ first asks $F'$ for its predecessor list, and then anonymously sends a ``lookup query'' to a random node $P_1$ in $F'$'s predecessor list. If any node in $P_1$'s successor list is closer to the ideal finger ID than $F$', $X$ detects $Y$ manipulated its fingertable. }
\label{fig:finger_checking}
\end{figure}

In particular, we let each node keep a small number of received fingertables (e.g., from lookups, secret neighbor surveillance, or random walks). From time to time, node $X$ chooses a random finger from one of the kept fingertables, say the $i$-th finger $F'$ of node $Y$, and asks $F'$ for its predecessor list (see Figure \ref{fig:finger_checking}). Then, after waiting a short random period of time, $X$ anonymously sends a ``lookup query'' to a random predecessor of $F'$ (say $P_1'$), and checks if any node in $P_1'$'s successor list is closer to the ideal finger ID than $F'$ (i.e., $F'$ is not the true finger $F$).

The intuition behind this is that if $Y$ replaces a honest finger $F$ with a malicious node $F'$, at least one of $F'$'s (true) predecessors should be closer to the ideal finger ID than $F'$. Hence, if $F'$ provides $X$ with its true predecessor list, the fingertable manipulation will be detected. Therefore, $F'$ has to manipulate its predecessor list to ensure that all the ``predecessors'' are malicious, so that the selected predecessor $P_1'$ can collude with $F'$ by providing a manipulated successor list that is consistent with the predecessor list provided by $F'$. On the other hand, however, $P_1'$ cannot freely manipulate its successor list, since $P_1'$ is under surveillance by its neighbors (i.e., secret neighbor surveillance). Therefore, if the adversary tries to manipulate a single finger ($F \rightarrow F'$), she has to sacrifice at least one malicious node, either $P_1'$ or $F'$ and $Y$.

\subsubsection{Secure Finger Update}

We can invoke the secret finger surveillance to limit the finger pollution attack: when $X$ obtains the result (say $F'$) of the finger-update lookup, it asks $F'$ for its predecessor list, and chooses a random predecessor $P_1'$ of $F'$ to perform the same checks as in the secret finger surveillance to verify $F'$ is the true finger; $X$ uses $F'$ to update its fingertable only when $F'$ passes these checks.

%%%%%%%%%%%

\subsection{Security Evaluation}

We use the following metrics to evaluate our security mechanisms. 
\begin{itemize}
	\item {\it fraction of remaining malicious nodes},
	\item {\it false positive rate}, i.e., the chance that a honest node is judged as a malicious node,
	\item {\it false negative rate}, i.e., the chance that a malicious node is not identified when being tested by a node,
	\item {\it false alarm rate}, i.e., the chance that there is no node identified in a report sent to the CA.
\end{itemize}

These metrics represent different aspects of security properties. Reduction of malicious nodes shows {\it effectiveness}, false positive/negative rates represent {\it accuracy}, and false alarm rate indicates {\it efficiency}. 

\begin{figure*}[t]
\centering

\begin{tabular}{ccc}

\subfloat[Secret neighbor surveillance] {\includegraphics[height=3.0cm]{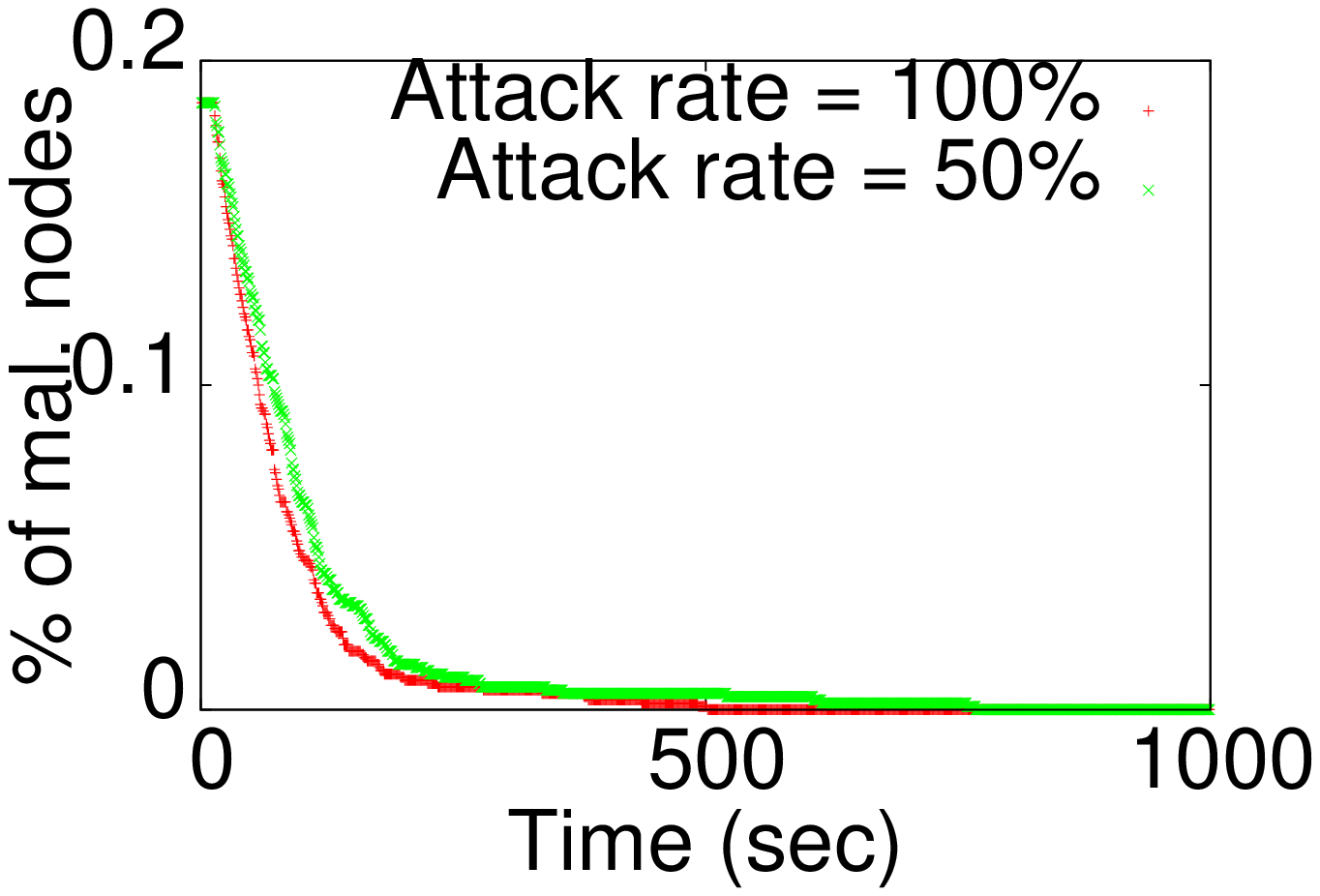} \label{fig:sl_manipulation} } 

   & \qquad \subfloat[Secret finger surveillance] {\includegraphics[height=3.0cm]{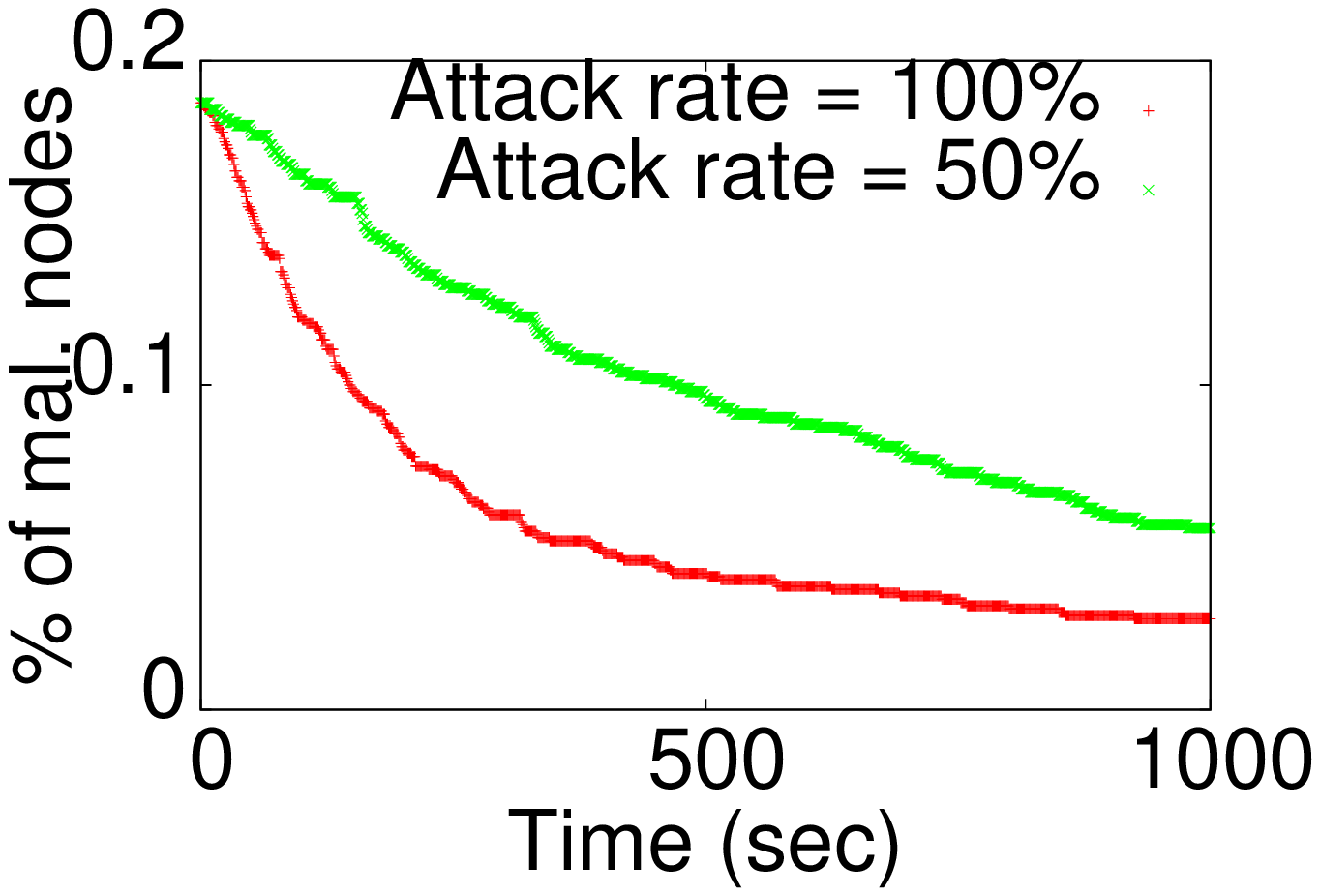} \label{fig:ft_manipulation} }
   
  & \qquad \subfloat[Secure finger update]{ \includegraphics[height=3.0cm]{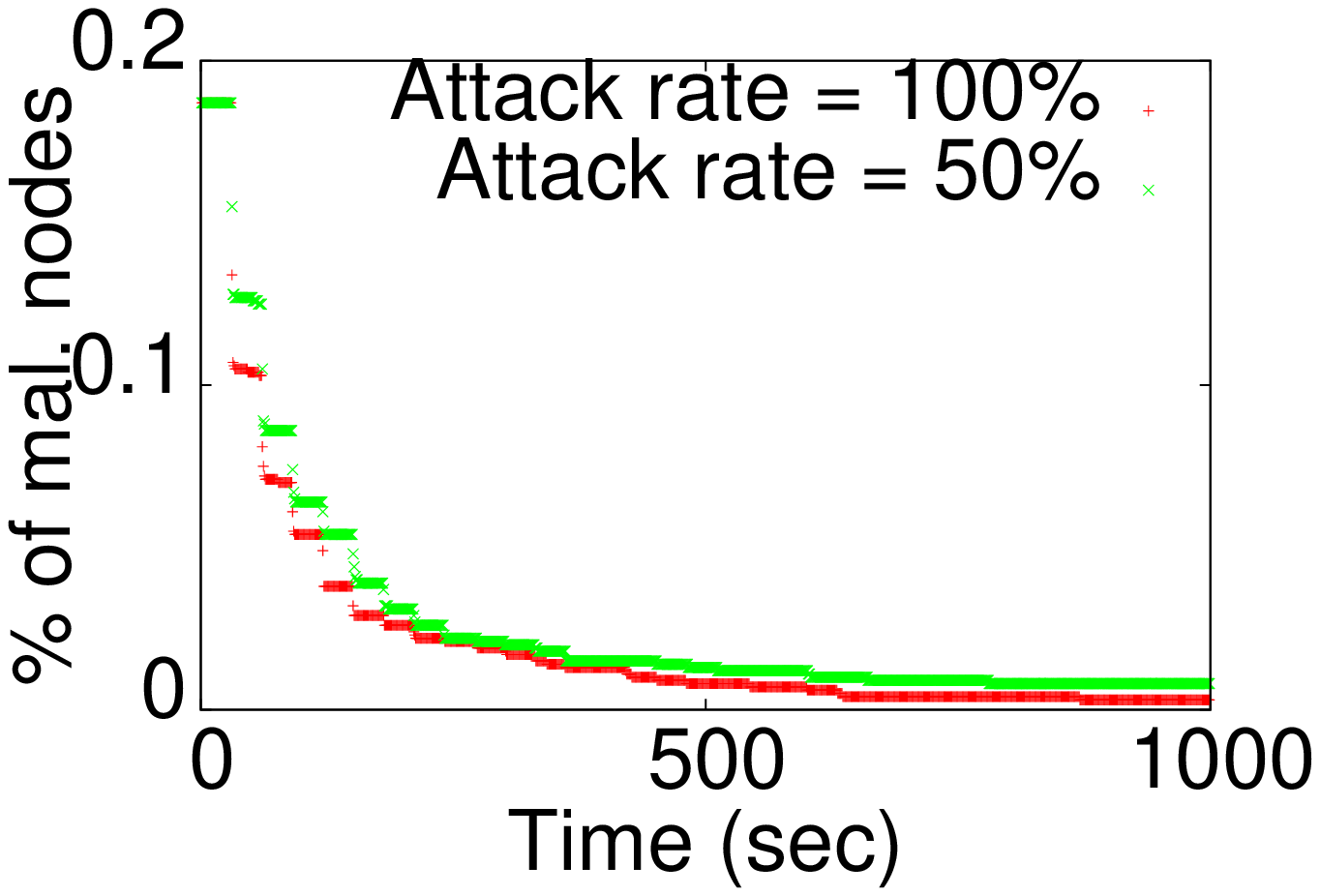} \label{fig:ft_pollution} } 
  
\end{tabular}

\caption[Optional caption for list of figures]{Simulation results of the security mechanisms}
\label{fig:security_mechanisms}
\end{figure*}

\subsubsection{Experiment Setup}
\label{sssec:exp_setup}

We developed an event-based simulator in C++ with about 3.0 KLOC. We consider a WAN setting, where latencies between each pair of peers are estimated using the King dataset~\cite{king}. We model node churn/join as an exponential distribution process $f(x) = \lambda e^{-(1/\lambda) x}$ with mean life time $\lambda$ minutes. We generate random network topologies of size $N=1000$ with 20\% malicious nodes. Each node maintains 12 fingers and 6 successors/predecessors (on the order of $\Theta(\log N)$). With similar configurations as related works~\cite{shadowwalker, chord}, we let each node run successor and predecessor stabilization protocols every 2 seconds, and performs finger update every 30 seconds. To discover malicious nodes, each peer performs security checks of secret neighbor surveillance and secret finger surveillance every 60 seconds\footnote{This is based on the frequency of stabilization and finger update as well as the node churn rate, and we found that doing security checks every 60s is sufficient to rapidly discover malicious nodes.}. To ensure high identification accuracy, each node keeps 6 latest received successor lists as proofs. Besides, we let each node perform one lookup every minute (we choose 1 min only because we want to test a large number of lookups within a relatively short simulation time).

\subsubsection{Experimental Results}

We see from Figure~\ref{fig:sl_manipulation} that the secret neighbor surveillance mechanism can rapidly identify malicious nodes that try to bias lookups. After a short time (20 mins), almost all malicious nodes are discovered. In comparison, the speed of discovering malicious nodes by the secret finger surveillance mechanism is relatively slower (as shown in Figure~\ref{fig:ft_manipulation}), but still it can identify over 80\% malicious nodes within 30 mins. The speed of identifying malicious nodes by the secure finger update mechanism is faster than that of the secure finger surveillance (as shown in Figure~\ref{fig:ft_pollution}), because the former is performed more frequently (at each finger update) and some malicious fingers contained in the successor list can also be detected by the secret neighbor surveillance.

\begin{table*}[t]\centering
	\caption{\small False positive/negative/alarm rates of the security mechanisms. $\lambda$ is mean life time of each node (in minute). Attack rate is 100\%. In fingertable manipulation/pollution attacks, checked malicious predecessors provide manipulated successor lists with 50\% chance.}
	\begin{tabular}{|c|c|c|c|c|c|c|}
	\hline
	\multirow{2}{*}{Security mechanisms}&	\multicolumn{2}{|c|}{False Positive}	&	\multicolumn{2}{|c|}{False Negative}	&	\multicolumn{2}{|c|}{False Alarm}	\\
	\cline{2-7}
				& 	$\lambda = 60m$	& $\lambda = 10m$	&	$\lambda = 60m$	& $\lambda = 10m$	&	$\lambda = 60m$	& $\lambda = 10m$	\\
	\hline	\hline
	Secret neighbor surveillance		&	0		&	0		&	0		&	0.52\%		&	0		&	0.52\%		\\
	\hline
	Secret finger surveillance &	0		&	0		&	14.02\%		&	19.55\%		&	0.18\%		&	1.55\%		\\
	\hline
	Secure finger update	&	0		&	0		&	14.08\%		&	18.48\%		&	0.33\%		& 	2.18\%		\\
	\hline
	\end{tabular}
	\label{tab:identification}
\end{table*}

The accuracy of our attacker discovery mechanisms is shown in Table \ref{tab:identification}. The false positive rate is 0 for all the three mechanisms even when the churn rate is very high (e.g., the mean life time for each node is 10 mins). This ensures that honest nodes will not be judged as malicious nodes by mistake. In addition, the secret neighbor surveillance has very low false negative rate (less than 0.6\%), which implies that any malicious nodes that try to bias lookups can be caught with high probability. 
We also see that the false negative rates for the secret finger surveillance and secure finger update mechanisms are relatively higher. This is because a malicious finger can pass the security checks if the randomly selected predecessor happens to be a colluding node and provides a successor list consistent with the malicious finger. However, over time, a malicious node can be identified with very high probability as shown in Figure~\ref{fig:security_mechanisms}.

\begin{figure}[h]
\centering
\includegraphics[height=3.5cm]{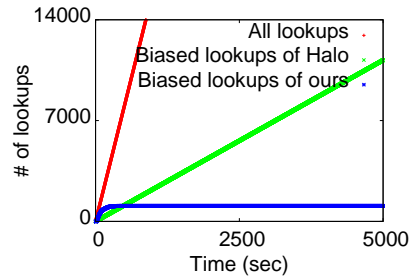}
\caption{Security comparison}
\label{fig:lookup_bias} 
\end{figure}

We also compare our scheme with a state-of-the-art secure DHT scheme Halo~\cite{Halo} in terms of the number of biased lookups over time. We calculate the ratio of biased lookups of Halo according to their analysis results~\cite{Halo} \S4.1 using the parameter $l=7$ as they suggested.  We can see from Figure~\ref{fig:lookup_bias} that after a short period of time, there are no more biased lookups in Octopus, while the number of biased lookups of Halo keeps increasing in linear with the total number of lookups. This demonstrates that our security mechanisms can fundamentally thwart active attacks.

\begin{figure}[h]
\centering
\includegraphics[height=3.5cm]{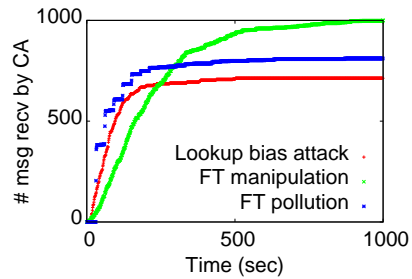}
\caption{The CA's workload}
\label{fig:CA_workload}
\end{figure}

Finally, We evaluate the workload of the CA in terms of the number of messages (including reports, proofs, and etc) processed over time.  We can see from Figure \ref{fig:CA_workload} that even during the peak time (the first 10 min), the CA only needs to process about 2 messages per second on average, which can be handled by most Internet servers.

\subsection{Other Attacks, Countermeasures, and Discussion}

We also consider other potential attacks, such as the selective denial-of-service attack and the relay exhaustion attack; we discuss these attacks and propose countermeasures in Appendix.

%% file: anonymous_lookup.tex
\section{Anonymity mechanisms of Octopus}
\label{sec:anony_lookup}

%In this section, we propose a secure and anonymous DHT lookup scheme (called Octopus) based on the security mechanisms described in Section~\ref{sec:secure_lookup}. We first motivate our design by describing the challenges.

\subsection{Problem Description}
\label{ssec:anony_problem}

\begin{figure}[h]
	\centering
	\includegraphics[height=4.0cm] {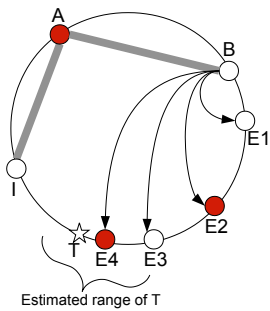}
	\caption{\small $T$ is the lookup target, $I \rightarrow A \rightarrow B$ is the anonymous path,  and $E_i$'s are queried nodes. }
\label{fig:one_path}
\end{figure}

In vanilla DHT systems, since the lookup initiator $I$ queries intermediate nodes directly, the queried nodes can easily infer the $I$'s identity.  A natural idea to let $I$ hide its identity by sending queries through an anonymous path, as in attacker identification mechanisms. An illustration of this is shown in Figure~\ref{fig:one_path}.

However, we note that {\it a single anonymous path is insufficient to achieve high levels of anonymity}. We use the example in Figure~\ref{fig:one_path} to show this. Assume queried nodes $E_2$ and $E_4$ are malicious, and the first relay $A$ is also malicious. With a single anonymous path, the adversary can learn that $E_2$ and $E_4$ belong to the same lookup since they are contacted by the same exit node $B$. Wang et al.~\cite{qiyan10} have shown that based on the positions of a few queried malicious nodes in the lookup, the adversary can narrow the range of the lookup target into a small set of nodes (called {\it range estimation attack}). Suppose there are $c$ concurrent lookups each having an estimation range of $d$ nodes; then, the adversary can know that $I$ is doing an lookup and its target is one of the $c\cdot d$ nodes.

\subsection{Anonymity Mechanisms}

\begin{figure}[h]
	\centering
	\includegraphics[height=3.0cm] {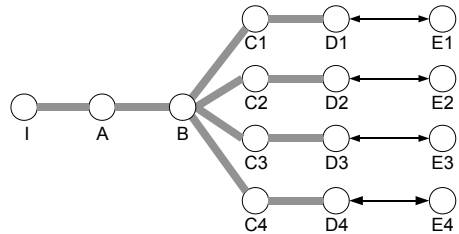}
	\caption{\small The structure of multiple anonymous paths in Octopus. $A$, $B$, $C_i$ and $D_i$ are relays.}
\label{fig:multi_paths}
\end{figure}

To address the limitation of a single anonymous path, we propose to split lookup queries over multiple anonymous paths, as shown in Figure~\ref{fig:multi_paths}. 

Using separate anonymous paths for different queries effectively disassociates the adversary's observations. The adversary only sees disjoint events from different concurrent lookups, but is unable to group queries belonging to the same lookup; in this case, it is much harder to apply the range estimation attack, thus substantially limiting the information leak. 

Moreover, to further blur the adversary's observations and make the range estimation attack even harder, we propose to add {\it dummy queries} in the lookup. Then, even though in rare cases the adversary can link two queried nodes in the same lookup (e.g., when all the relays used for the two queries are malicious), the adversary is unable to tell whether they are dummy queries or true queries, and the result of the range estimation attack would be incorrect with a dummy query.  

We note that using multiple anonymous paths is important to ensure effectiveness of dummy queries, because in the single-anonymous-path scenario, observed queries are linkable due to the common exit relay and hence dummy queries can be distinguished based on the positions of observed queries. In comparison, with multiple anonymous paths, identifying dummy queries is much harder.

\subsection{Anonymity Evaluation}
\label{ssec:anonymity_evaluation}

We analyze the best strategies for an adversary to infer the lookup target $T$ and the initiator $I$ based its observations, and calculate the target anonymity $H(T)$ and the initiator anonymity $H(I)$. We use entropy to quantify $H(T)$ and $H(I)$. We let $O$ denote the set of possible observations of the adversary (including null observation). To measure the system as a whole, we have:   
\begin{equation} 
	H(T) = \sum_{o \in O} P(o) \cdot H(T|o) , 
	H(I) = \sum_{o \in O} P(o) \cdot H(I|o)  \label{equ:entropy_def}	
\end{equation} 
where $P(o)$ is the probability of observation $o$ occurring. 

To calculate the maximum information leak, we make the following assumptions in the anonymity calculation. First, we assume the network is static, since network dynamics can obscure the adversary's observations and make it more difficult to extract information about the initiator/target. Second, we only consider passive attacks, as active attackers can be quickly identified by our security mechanisms and consequently the adversary will lose observers to carry out passive attacks.

\subsubsection{The Adversary's Observations}
\label{sssec:observation}

An observation $o$ consists of a (large) number of observed events, which are message transmissions seen by malicious nodes. Each observed event can provide information, such as sender/receiver IDs, message content, and transmission time. The adversary can log all the observed events, and try to derive useful information from any combinations of them.

{\bf Observations of queries.} There are two cases where a query is observed (i.e., the queried node is identified): 1) the queried node itself is malicious, or 2) the exit relay is malicious. The adversary can link an observed query backwards to $I$ in two ways. One is through direct connection of compromised relays on the anonymous path. For example, if the relays $A$ and $C_i$ and the queried node $E_i$ are malicious (refer to Figure \ref{fig:multi_paths}), $E_i$ is observed and can be linked to $I$ using $C_i$ and $A$ as bridges. The other is through linkability of relays to $I$ in the random walk. For instance, if $D_i$ is compromised and linkable to $I$ in the random walk, then $E_i$ is observed and linkable to $I$ using $D_i$ as a shortcut. It is possible to use both approaches at the same time. 
%One example is that $B$ and $D_i$ are compromised and $B$ is linkable to $I$ in the random walk; in this case, the adversary can first link $E_i$ to $B$ via $D_i$ and then further link it to $I$ through $B$. 

Furthermore, considering all queries in the same lookup together can help the adversary link more queries to $I$. Since all $(C_i, D_i)$ in the same lookup are connected to the same relay $B$, if there exists one query linkable to both $I$ and $B$, then any other queries that are linkable to $B$ can also be linked to $I$. In the rest paper, we use ``a linkable query'' to specifically refer to a query that is {\it observed} and {\it linkable} to $I$.

{\bf Observations of the initiator.} The adversary's goal is to link $I$ and $T$; knowing only one of them is useless to the adversary. Therefore, the pre-condition of compromising the target anonymity is to observe $I$. Since $I$ is directly connected to $A$, the adversary can observe $I$ as long as $A$ is malicious. In addition, $I$ is also observable in random walks; hence, another case for $I$ being observed is that there exists at least one malicious relay linkable to $I$ in a random walk.

{\bf Observation of the target.} For similar reasons, in order to compromise the initiator anonymity, the adversary has to know $T$. We note that $T$ is not necessarily contacted during the lookup, since the aim of the lookup is to find the IP address of $T$, which can be learnt from query replies of intermediate nodes (after the lookup is done, $T$ might be contacted due to some application needs, but we do not consider this as part of the lookup). Yet we assume that each node can tell whether itself is a target node based on its role in the application. For example, in DHT-based anonymous communication, a node can learn that itself is a lookup target if it is selected as a relay of an anonymous circuit. Therefore, we concern $T$ as observed if $T$ itself is a malicious node.

%%%%%%%%%%%%%%%%%

\subsubsection{Initiator Anonymity}
\label{sssec:initiator_anonymity}

To calculate the initiator anonymity, we divide the observations of the adversary into two categories. 
\begin{itemize}	 
	\item $o_n$: the observation occurring when $T$ is not observed	 
	\item $O_o$: the set of observations occurring when $T$ is observed	 
\end{itemize}
According to~\ref{equ:entropy_def}, $H(I)$ is calculated as:
 
\begin{equation}
	H(I) = P(o_n) \cdot H(I|o_n) + \sum_{o_o \in O_o} P(o_o) \cdot H(I|o_o)
\end{equation}

When $T$ is not observed, the entropy of $I$ is maximized. Presuming that the adversary can exclude malicious nodes from the anonymity set of $I$, we have:
 
\begin{equation}
	H(I|o_n) = \log_2 \left( (1-f)\cdot N \right)
\end{equation}

Let $\mathcal{R}^l_T$ denote the set of non-dummy queries linkable to $I$ in the lookup whose target is $T$. Then, based on whether $\mathcal{R}^l_T$ is an empty set, we calculate $H(I|o_o)$ as follows:
 
\begin{eqnarray}
	H(I|o_o) &=& P(\mathcal{R}^l_T = \emptyset) \cdot H'(I|o_o)  \nonumber \\
		  &+& (1 - P(\mathcal{R}^l_T = \emptyset)) \cdot H''(I|o_o)
\end{eqnarray}

When there is no linkable non-dummy query, $I$ is unlinkable with $T$. However, since some of the initiators of concurrent lookups can be observed by the adversary, we have:
 
\begin{eqnarray}
	H'(I|o_o) &=& P(I\_obsv) \cdot \log_2 (\#obsv\_hon\_init) \nonumber \\
	     &+& (1 - P(I\_obsv)) \cdot \log_2 ((1-f) \cdot N)
\end{eqnarray}

Let $\Psi^l$ denote the set of concurrent lookups that have at least one linkable query, and $\psi_T$ denote the lookup with target $T$. When $\mathcal{R}^l_T \neq \emptyset$, $\psi_T \in \Psi^l$. 
Each lookup in $\Psi^l$ is possible to be $\psi_T$. Therefore, we have:
 
\begin{equation}
	H''(I|o_o) = - \sum_{\psi \in \Psi^l} P(\psi = \psi_T |o_o) \cdot \log_2 P(\psi = \psi_T |o_o)
\end{equation}

Because the density of queries close to the target is higher than other regions on the ring, for $\psi_T$ it is highly likely that the last queried node in $\mathcal{R}^l_T$ is located very close to $T$. Therefore, the adversary can assign probability to each candidate initiator based on the minimum distance (i.e. number of hops) between its queried nodes and $T$. In particular, let $\mathcal{Q}^l_{\psi}$ denote the set of linkable queries in $\psi$, $\psi \in \Psi^l$, and let $\xi(x)$ denote the probability that for $\psi_T$ the minimum distance from linkable queried nodes to $T$ is $x$. $\xi(x)$ can be obtained via pre-simulations of the lookup. Then, we can calculate $P(\psi = \psi_T | o_o)$ as follows: 
\begin{equation}
	P(\psi = \psi_T | o_o) \approx \frac{\xi(\min_{E \in \mathcal{Q}^l_{\psi}} dist(E, T))}{\sum_{\psi' \in \Psi^l} \xi(\min_{E' \in \mathcal{Q}^l_{\psi'}} dist(E', T))}
\end{equation}

%%%%%%%%%%%

\subsubsection{Target Anonymity}
\label{sssec:target_anonymity}

We categorize the adversary's observations into three classes: 
\begin{itemize}	 
	\item $o_n$: the observation occurring when $I$ is not observed  
	\item $O_l$: the set of observations occurring when there is at least one linkable query in the lookup  
	\item $O_d$: the set of observations occurring when there is no linkable query in the lookup	 
\end{itemize}
According to Equation (\ref{equ:entropy_def}), we have:  
\begin{eqnarray}
	H(T) &=& P(o_n) \cdot H(T|o_n) + \sum_{o_l \in O_l} P(o_l) \cdot H(T|o_l) \nonumber \\
	     &+& \sum_{o_d \in O_d} P(o_d) \cdot H(T|o_d)
\end{eqnarray}

Since the adversary has to know $I$ at the first place, the entropy of $T$ is maximum when $I$ is not observed, i.e. $H(T|o_n) = \log_2 N$.

{\bf Calculation of $H(T|o_l)$.} When there exist queries that are linkable to $I$, the adversary can adopt range estimation attack to narrow the range of $T$. The output of this attack is a lower bound and an upper bound of $T$'s location on the ring. 

We temporarily assume that all queries observed by the adversary are non-dummy queries (how to deal with dummy queries is shown later). Suppose there are two or more linkable queries in the lookup. Let $E_i$ and $E_j$ denote the first and the last linkable queried nodes, respectively. Since nodes succeeding $T$ will not be queried in the lookup, $E_j$ can be used as a lower bound of $T$. An upper bound of $T$ can be obtained based on the fact that the lookup always greedily queries the finger that is precedingly closest to the target. In particular, the adversary first decides the queried nodes between $E_i$ and $E_j$ by (locally) simulating the lookup from $E_i$ to $E_j$. The initial upper bound is set as $E_i$; then for each pair of consecutive queries ($E_k, E_{k+1}$) between $E_i$ and $E_j$, $i \leq k \leq j-1$, the adversary finds out the index of $E_{k+1}$ in $E_k$'s finger table (say $p$) and uses the $(p+1)$-th finger of $E_k$ to update the upper bound. 

Since the density of queries close to $T$ on the ring is higher than other regions, nodes located closer to $T$ in the estimation range are more likely to be $T$.  We let $\gamma(i, z)$ denote the probability that the $i$-th node (clockwise) in an estimation range of size $z$ is the target, $1 \leq i \leq z$. The probability distribution of $\gamma(i, z)$ can be obtained by pre-simulation of the lookup. 
Note that the range estimation attack is inapplicable when the lookup has only one linkable query (say $E_i$), but the adversary can use the successor of $E_i$ as the lower bound of $T$ and the predecessor of $E_i$ as the upper bound of $T$, and assign higher probabilities to the nodes closer to the lower bound in the estimation range. 

Now we discuss how to deal with dummy queries. Let $\mathcal{Q}^l_I$ denote the set of linkable queries in the lookup performed by $I$, and $\mathcal{R}^l_I$ denote the set of linkable non-dummy queries, $\mathcal{R}^l_I \subseteq \mathcal{Q}^l_I$. Based on whether $\mathcal{R}^l_I$ is an empty set, $H(T|o_l)$ can be calculated as: 
\begin{eqnarray}
	H(T|o_l) &=& P(\mathcal{R}^l_I=\emptyset) \cdot H_m \nonumber \\
	&+& (1 - P(\mathcal{R}^l_I=\emptyset)) \cdot H'(T|o_l)
\end{eqnarray}

$H_m$ denotes the entropy when all linkable queries are dummies. In this case, the linkable queries cannot provide any information about $T$. However, the adversary can observe all (concurrent) malicious target nodes, and $T$ has chance $f$ to be one of them. Therefore, we have: 
\begin{eqnarray} 
\label{equ:max_entropy}
	H_m &=& (1-f) \cdot \log_2 ((1-f)\cdot N) \nonumber \\
	&+& f \cdot \log_2 (\#mal\_targets) 
\end{eqnarray}

When $\mathcal{R}^l_I$ is non-empty, a range estimation attack based on $\mathcal{R}^l_I$ can produce a minimum range of $T$. Whereas, an estimation range calculated using any dummy query will be incorrect. Due to use of anonymous paths, an individual dummy query is indistinguishable from any non-dummy query. Nevertheless, the adversary can base on timing and location relationships between queries to filter out some subsets of $\mathcal{Q}^l_I$ that contain dummy queries. In particular, any subset of queries that violates the following rules must contain at least one dummy query:
\begin{itemize}  
	\item if $E_i$ is queried before $E_j$, then $E_i$ must precede $E_j$	 
	\item if $E_i$ and $E_j$ are the first and last queried nodes in the subset, then any other query must be on the path of the {\it virtual lookup} from $E_i$ to $E_j$	 
\end{itemize}

Note that the above approach cannot remove all subsets that contain dummy queries. Let $\mathcal{S}_I$ denote all subsets of $\mathcal{Q}^l_I$ that pass the above filtering test, $\mathcal{S}_I \subseteq 2^{\mathcal{Q}^l_I}$. Since all queries in $\mathcal{R}^l_I$ are non-dummy, $\mathcal{R}^l_I$ will pass the filtering test, i.e., $\mathcal{R}^l_I \in \mathcal{S}_I$. From the adversary's prospective, each element of $\mathcal{S}_I$ is possible to be $\mathcal{R}^l_I$. The best strategy for her is to assign different probability to each element in $\mathcal{S}_I$ according to the pre-calculated probability distribution of $\mathcal{R}^l_I$. We use two variables to characterize $\mathcal{R}^l_I$: the number of queries in $\mathcal{R}^l_I$, and the largest hop in the virtual lookup from the first query in $\mathcal{R}^l_I$ to the last query.\footnote{The largest hop means the largest ID difference between two consecutively queried nodes. The largest hop also implies the number of hops in the lookup. These two characteristics are a close approximation of the adversary's observation on $\mathcal{R}^l_I$.} 

Let $X$ denote an arbitrary node that is contained in any estimation range. Then:  
\begin{equation}
	H'(T|o_l) = - \sum_X P(X=T|o_l) \cdot \log_2 P(X=T|o_l)
\end{equation}

Let $G(s)$ denote the estimation range computed based on $s$, $s \in \mathcal{S}_I$. Let $loc(G(s), X)$ denote the location of $X$ in the estimation range $G(s)$, and $|\cdot|$ denote the number of elements of a set. Then, we have: 
\begin{equation}
	P(X=T|o_l) = \sum_{s \in \mathcal{S}_I} P(s = \mathcal{R}^l_I | o_l) \cdot \gamma(loc(G(s), X), |G(s)|)
\end{equation}

Let $V(s)$ denote the largest hop in the virtual lookup based on $s$, and $\chi(x, y)$ denote the probability that a set of $x$ queries with the largest hop in the virtual lookup being $y$ is $\mathcal{R}^l_I$. Then, we have:
 
\begin{equation} 
\label{equ:real_set}
	P(s = \mathcal{R}^l_I|o_l) \approx \frac{\chi(|s|, V(s))}{\sum_{s' \in \mathcal{S}_I} \chi(|s'|, V(s'))}
\end{equation}
$\chi(x,y)$ is obtained by pre-simulations of the lookup.

{\bf Calculation of $H(T|o_d)$.} 
There are three possible cases when there is no linkable query:
\begin{itemize}	 
	\item $case_1$: there is no query observed by the adversary	 
	\item $case_2$: there is at least one (observed) query that is linkable to $B$	 
	\item $case_3$: there is no query linkable to $B$ but at least one query is observed by the adversary	 
\end{itemize}
Let $H_1$, $H_2$, and $H_3$ denote the entropy of $T$ in the three cases, respectively. Then, we have: 
\begin{eqnarray}
	H(T|o_d) &=& P(case_1) \cdot H_1 + P(case_2) \cdot H_2 \nonumber \\
	&+& P(case_3) \cdot H_3
\end{eqnarray}

In the first case, since no information is learnt from queries, $H_1$ is calculated as Equation (\ref{equ:max_entropy}). 

For the second case, although $I$ is disassociated with any observed queries, the adversary can group queries belonging to the same lookup based on whether they are linkable to a common relay $B$. Furthermore, she can calculate estimation ranges for each concurrent lookup that contains queries linkable to $B$, and consider all estimation ranges as possible candidates for the true estimation range of $T$. 

In particular, we let $\mathcal{R}^B_I$ denote the set of non-dummy queries linkable to $B$ in the lookup performed by $I$, and $H_2$ can be calculated as follows. 
\begin{equation}
	H_2 = P(\mathcal{R}^B_I = \emptyset) \cdot H_m + (1 - P(\mathcal{R}^B_I = \emptyset)) \cdot H_2'
\end{equation}
where $H_2'$ denotes the entropy when there is at least one non-dummy query linkable to $B$. If $T$ is malicious, the adversary can reduce the candidates of $T$ down to the set of observed malicious targets; otherwise, she needs to rely on the queries linkable to $B$ to infer $T$. Therefore, we have: 
\begin{eqnarray}
	H_2' &=& f \cdot \log_2 (\#mal\_targets)  \\
	     &-& (1-f) \cdot \sum_X P(X=T|o_d) \cdot \log_2 P(X=T|o_d) \nonumber 
\end{eqnarray}

Let $\Psi^B$ denote the set of concurrent lookups that have at least one query linkable to $B$, and $\psi_I$ denote the lookup performed by $I$. Let $\mathcal{R}^B_{\psi}$ denote the set of non-dummy queries linkable to $B$ in $\psi$, $\psi \in \Psi^B$, and let $\mathcal{S}_{\psi}$ denote all subsets of queries (linked to $B$) in $\psi$ that pass the filtering test. Then, we have:
 
\begin{eqnarray}
	 && P(X=T|o_d) = \sum_{\psi \in \Psi^B} P(\psi = \psi_I|o_d) \\
	&& \quad \cdot \sum_{s \in \mathcal{S}_{\psi}} P(s = \mathcal{R}^B_{\psi}|o_d) \cdot \gamma(loc(G(s), X), |G(s)|)  \nonumber 
\end{eqnarray}

Since $I$ is unlinkable to any observed queries, each concurrent lookup is equally likely to be $\psi_I$. We have: 
\begin{equation}
	P(\psi = \psi_I|o_d) = \frac{1}{|\Psi^B|}, \quad \forall \psi \in \Psi^B
\end{equation}
$P(s=\mathcal{R}^l_{\psi}|o_d)$ is calculated the same as Equation (\ref{equ:real_set}).

In the last case, since all observed queries are disassociated with each other, the range estimation attack cannot be applied. Let $\mathcal{R}^o_I$ denote the set of observed non-dummy queries in the lookup performed by $I$. Similar to the above cases, we have:  
\begin{eqnarray}
	H_3  &=& P(\mathcal{R}^o_I = \emptyset) \cdot H_m + (1 - P(\mathcal{R}^o_I = \emptyset)) \cdot H_3' \\
	H_3' &=& f\cdot \log_2(\#mal\_targets) \\
	     &-& (1-f) \cdot \sum_X P(X=T|o_d) \cdot \log_2 P(X=T|o_d) \nonumber	
\end{eqnarray}

Let $E_I$ denote the query closest to $T$ in $\mathcal{R}^o_I$. Then, the adversary can use $E_I$'s successor as the lower bound of the estimation range of $T$ and $E_I$'s predecessor as the upper bound, and assign probabilities to the nodes in the range according to a pre-calculated probability distribution of $T$. Let $\mathcal{Q}^o$ denote the set of observed queries of all concurrent lookups, and $G(E)$ denote the estimation range based on $E$. Then, we have:  
\begin{eqnarray}
	P(X = T|o_d) &=& \sum_{E \in \mathcal{Q}^o} P(E = E_I | o_d) \\
	&& \cdot \gamma(loc(G(E), X), N-1) \nonumber
\end{eqnarray}  

Based on the observations, the adversary is unable to tell which query is more likely to be $E_I$. Therefore, we have: 
\begin{equation}
	P(E = E_I|o_d) = \frac{1}{|\mathcal{Q}^o|}, \quad \forall E \in \mathcal{Q}^o
\end{equation}

\subsection{Results and Comparisons}

\begin{figure*}[t]
\centering

\begin{tabular}{cc}

\subfloat[Initiator anonymity.] {\includegraphics[height=3.5cm]{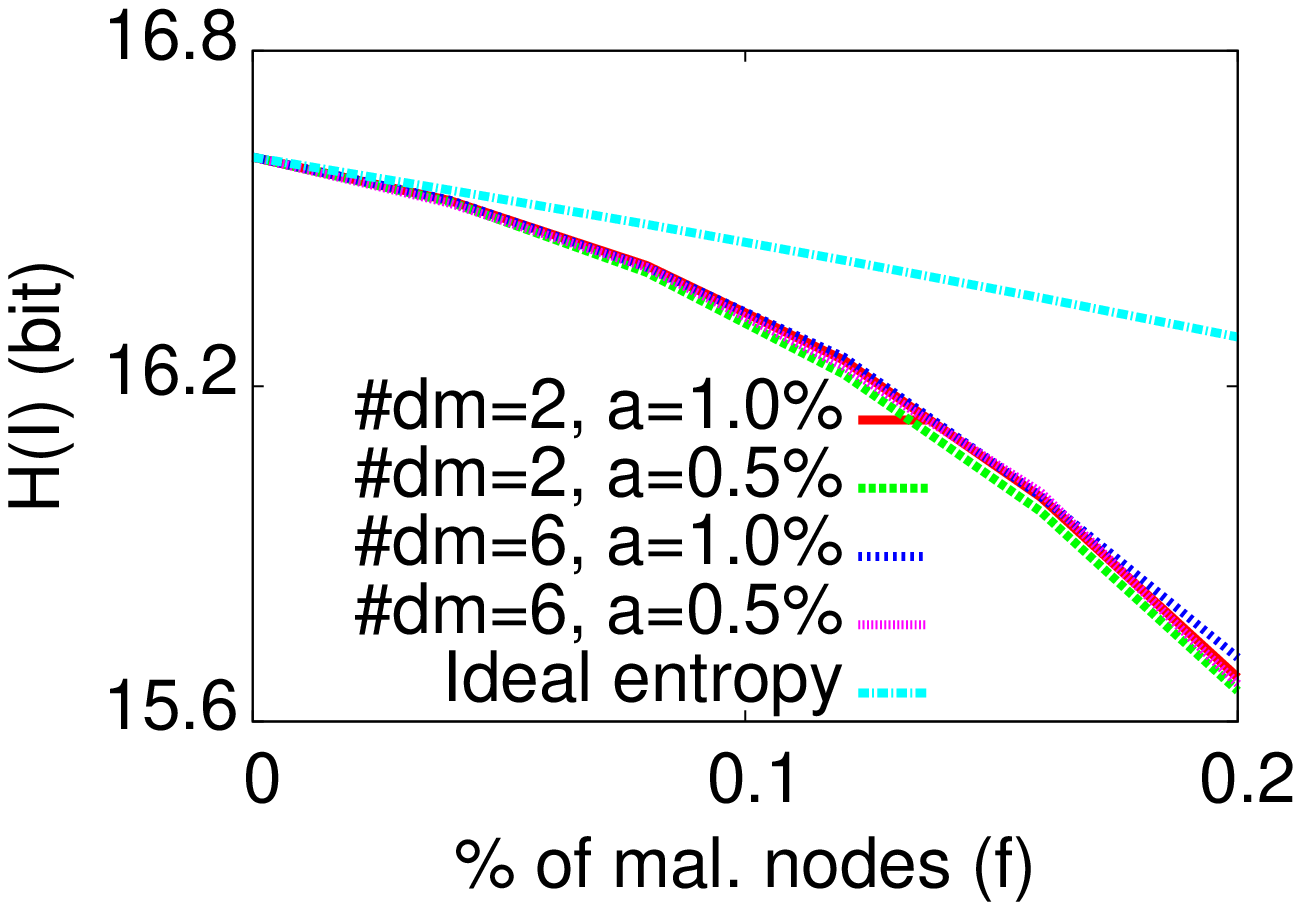} \label{fig:init_anonymity} } 

  & \qquad \qquad \subfloat[Target anonymity.]{ \includegraphics[height=3.5cm]{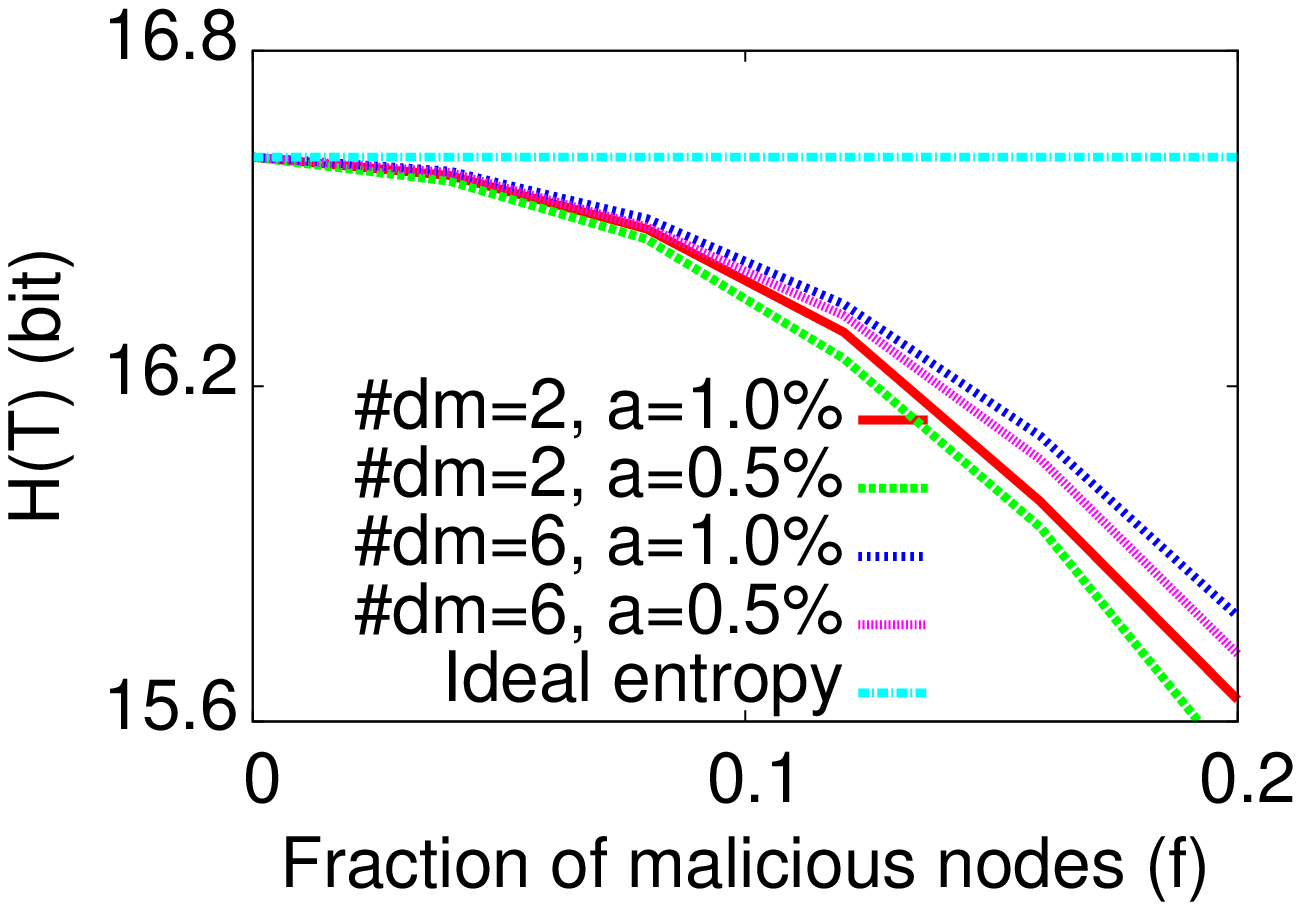} \label{fig:target_anonymity} } 

\end{tabular}

\caption[Optional caption for list of figures]{Anonymity evaluation of Octopus. $a$ is the concurrent lookup rate.}
\label{fig:anonymity}
\end{figure*}

\begin{figure*}[t]
\centering

\begin{tabular}{cc}

 \subfloat[Initiator anonymity.] {\includegraphics[height=3.5cm]{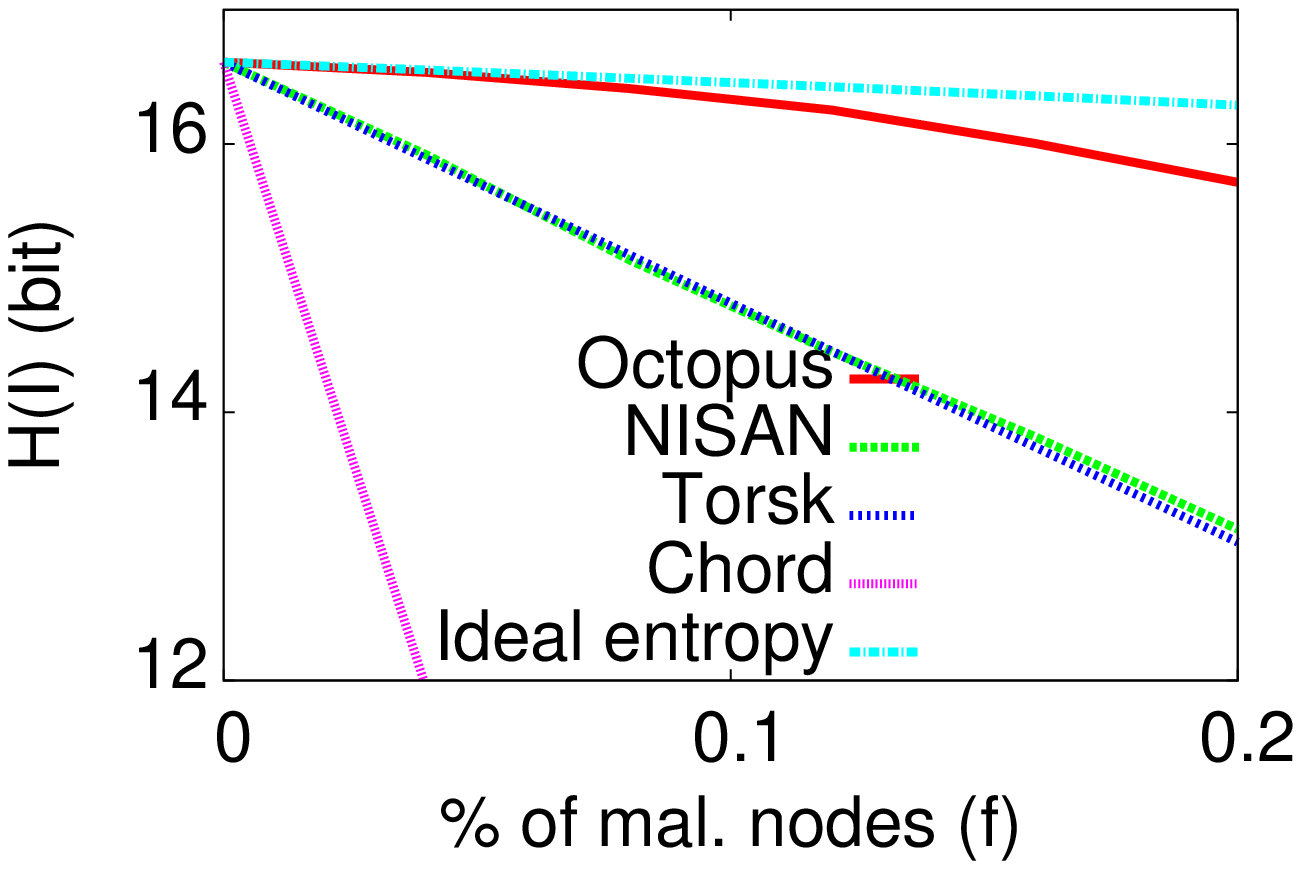} \label{fig:init_comparison} }
 
  & \qquad \qquad \subfloat[Target anonymity.]{ \includegraphics[height=3.5cm]{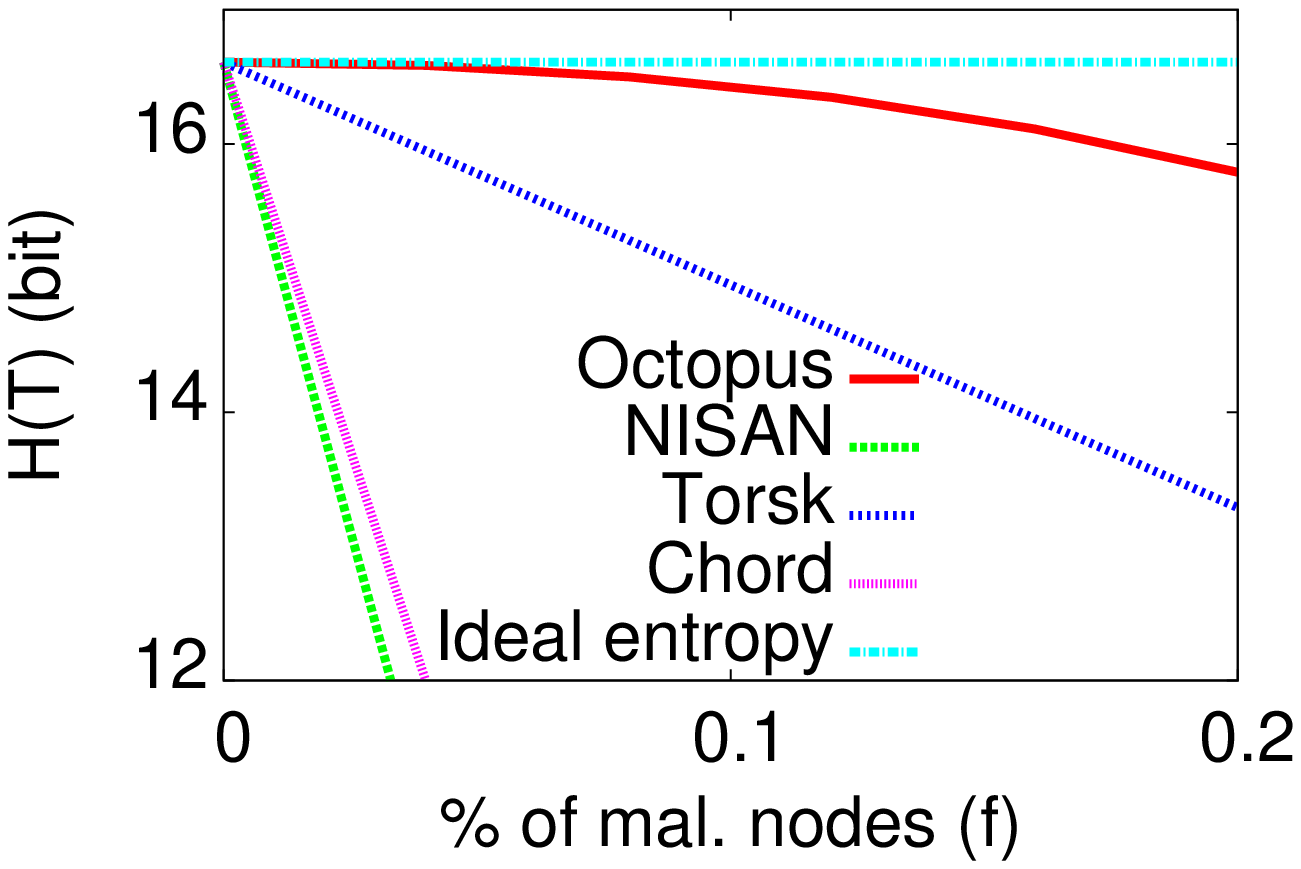} \label{fig:target_comparison}} 
  
\end{tabular}

\caption[Optional caption for list of figures]{Anonymity comparison. $a=1\%$.}
\label{fig:anonymity}
\end{figure*}

We developed a simulator for anonymity measurements in C++ with about 1.3 KLOC. The results are shown in Figure~\ref{fig:anonymity}.  With network size $N=100\,000$, concurrent lookup rate $a = 1\%$, $f=20\%$ malicious nodes, and 6 dummies, Octopus only leaks 0.57 bits of information about the initiator and 0.82 bits of information about the target. We compare Octopus with the base-line scheme Chord~\cite{chord} and the state-of-the-art anonymous DHT lookups NISAN~\cite{NISAN} and Torsk~\cite{torsk} (we do not explicitly compare Octopus with secure DHTs such as Halo~\cite{Halo}, since they leak more information than Chord.). We can see that in the same setting, NISAN and Torsk leak about 3.3 bits of information about the initiator, which is about 6 times more than Octopus. As for the target anonymity, the information leak for NISAN and Torsk is 11.3 bits and 3.4 bits, which is 13 times and 4 times more than that of Octopus, respectively.

%% file: experiment.tex
\section{Performance Evaluation}
\label{sec:performance}

%In this section, we evaluate the performance of Octopus in terms of bandwidth overhead and lookup latency, and compare it against vanilla Chord lookup.

\begin{table}[t]\centering
	\caption{\small Performance comparison. $\tau$ is the time interval between two consecutive lookups.} 
	\begin{tabular}{|c|c|c|c|c|}
	\hline
	\multirow{2}{*}{Schemes}&	\multicolumn{2}{|c|}{Lookup Latency (sec)}	&	\multicolumn{2}{|c|}{Bandwidth Consumption (kbps)}		\\
	\cline{2-5}
						& 	Mean	& Median	&	$\tau=5min$	& $\tau=10min$	\\
	\hline	\hline
	Octopus		&	2.15		&	1.61		&	5.91						&	4.30			\\
	\hline
	Chord~\cite{chord} 			&	1.35		&	0.35		&	0.29						&	0.28			\\
	\hline
	Halo~\cite{Halo}				&	6.89		&	1.79		&	0.71						&	0.37			\\
	\hline
	\end{tabular}
	\label{tab:efficiency}
\end{table}

\subsection{Lookup Latency}

Lookup latency is one of the most important performance factors for DHT systems.  We measure the lookup latency of Octopus using PlanetLab with 207 randomly selected nodes. We use boost C++ library\footnote{www.boost.org} (mainly UDP asynchronous read/write of Boost.Asio) to build the communication substrate. We let each node perform 2000 lookups independently using randomly picked lookup keys. For each lookup, we record the latency from the time of sending out the first query till the time of receiving the lookup result. 

\begin{figure}[h]
	\centering
	\includegraphics[height=3.2cm, width = 5cm] {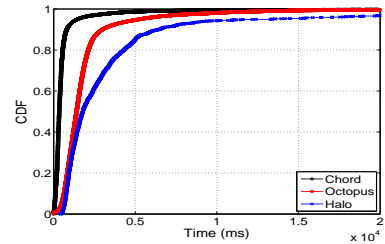}
	\caption{\small Comparison of lookup latency on Planetlab.}
\label{fig:lookup_time}
\end{figure}

For comparison, we use the same methodology to implement Chord~\cite{chord} and Halo~\cite{Halo}, and measure their lookup latencies in the same network environment. We choose Halo because it is one of the state-of-the-art secure DHT lookup schemes and it is also based on Chord overlay. For Halo, we use degree-2 recursion with redundant parameter $8\times4$, as suggested in their paper~\cite{Halo} to provide fairly strong security guarantee. The experimental results are presented in Figure \ref{fig:lookup_time} and Table \ref{tab:efficiency}. We can see that while the lookup latency of Octopus is relatively longer than that of Chord due to more transmission for security and anonymity needs, it is smaller than that of Halo, which only provides security guarantees. The outperformance is because Octopus does not rely on redundant lookups, while in Halo a lookup is not completed until all redundant lookups' results are returned.  

\subsection{Bandwidth Overhead}

We also compare Octopus with Chord and Halo in terms of bandwidth cost. We adopt the same configuration as described in Section \ref{sssec:exp_setup} for each of the DHT lookups, and consider an overlay network with 1\,000\,000 nodes\footnote{We use the following parameters to estimate the bandwidth overhead. Each routing state item (such as fingers or successors) is 10 bytes. We use ECDSA signature (40 bytes) for authentication with a 4-byte timestamp, and AES-128 for onion encryption. Each certificate is 50 bytes, including the node's IP address (6 bytes), the node's public key (20 bytes), expire time (4 bytes), and the CA's signature (20 bytes).}. We can see from Table~\ref{tab:efficiency} that Octopus does incur higher communication overhead than Chord and Halo, in order to achieve high levels of anonymity; however, the bandwidth cost of Octopus is still reasonable (only a few kbps), which is affordable even for low-end clients with limited bandwidth.

%% file: background.tex
\section{Related Work}
\label{sec:related_work}

\subsection{Secure DHT Lookups}

A major school of proposals to securing DHT lookups uses {\it redundancy}.   Castro et al.~\cite{castro} proposed a robust DHT system that relies on redundant lookups. Each key is replicated among several replica nodes (typically the neighbors of the key owner). Instead of doing a single lookup, the initiator performs multiple redundant lookups towards all the replicas. The lookup result would be correct as long as one of the redundant lookups is not biased. The limitation of this approach is that the redundant lookups tend to converge to a small number of nodes close to the target, and one malicious node in this set could infect many redundant lookups. Much subsequent work (such as~\cite{Cyclone, salsa, Halo}) focuses on disentangling the redundant lookup paths to provide better security. Cyclone~\cite{Cyclone} partitions nodes into $r$ Chord sub-rings based on similarity of node IDs, and has $r$ redundant lookups routed through the $r$ sub-rings independently. Salsa~\cite{salsa} uses a new virtual-tree-based DHT structure, in which any two nodes share few global contacts so that redundant messages can proceed along different paths. Halo~\cite{Halo} does not change the underlying DHT structures, but uses the original Chord overlay and performs redundant searches towards {\it knuckles} -- nodes that have fingers pointing to the target.

While effective in ensuring security, these redundant-lookup-based approaches are incapable of preserving anonymity, since redundant transmission creates opportunities for an adversary to gain information about the lookup initiator and/or target~\cite{prateek08}. ShadowWalker~\cite{shadowwalker} embeds redundancy into the DHT itself and uses {\it shadows} (nodes in redundant topologies) to verify each step of a lookup. Unfortunately, Schuchard et al.~\cite{eclipse_attack} found that ShadowWalker is vulnerable to {\it eclipse attack}, where the entire set of shadows of a certain node are compromised, leading to other nodes' routing states being infected.  They also showed that increasing the dimension of redundant topologies can mitigate the eclipse attack, but the resultant performance cost is prohibitively high.

Another major school of research on secure DHT lookups leverages cryptographic techniques. Mymric~\cite{Myrmic} uses an online certificate authority to sign each node's routing state. The major limitation of Myrmic is that for each node join/churn, the central authority has to update the certificates for all related nodes. Young et al.~\cite{RCP} proposed two schemes RCP-I and RCP-II that use threshold signature and distributed key generation to avoid the reliance of a central authority. In their schemes, the verification information on each message is collaboratively generated by a threshold number of nodes, rather by a central authority. 

All these secure DHT lookup schemes are not designed to preserve anonymity. Lookup keys are revealed during queries, and identities of lookup initiators are easily exposed due to directly contacting intermediate nodes. 

\subsection{Secure and Anonymous DHT Lookups}
NISAN~\cite{NISAN} is among the first to try to provide both security and anonymity guarantees in DHT systems. For security purpose, each queried node is required to provide its entire fingertable, so that the lookup initiator can apply {\it bound checking} on it to limit manipulation of fingertables. NISAN also uses redundancy to enhance security. The authors proposed a greed-search mechanism to query multiple nodes at each step and combine the query results to tolerate misinformation. On the other hand, acquiring the entire fingertable also helps protect the anonymity of lookup targets, since the lookup keys are not revealed to intermediate nodes. Nevertheless, NISAN can only provide very limited anonymity protection. Wang et al.~\cite{qiyan10} showed that a passive adversary is able to narrow the range of a lookup target down to a small number of nodes, by analyzing the locations of observed queries (called {\it range estimation attack}). %In addition, similar to vanilla and security-only DHT lookups, NISAN does not preserve initiator anonymity.

Torsk~\cite{torsk} is a DHT-based anonymous communication system. A key component of Torsk is a proxy-based anonymous DHT lookup. The idea is that a lookup initiator performs a random walk on the overlay to find a random node (called {\it buddy}), and requests the buddy to perform the lookup on its behalf. Because Torsk uses Myrmic~\cite{Myrmic} to secure lookups, it has the same limitation as Myrmic -- requiring an online central authority to sign each node's routing state. In addition, as we analyzed in Section \ref{ssec:anony_problem}, a single proxy structure is insufficient to provide high levels of anonymity: the information learnt by the range estimation attack can be used to launch {\it relay exhaustion attack}~\cite{qiyan10}. 

Recently, Backes et al.~\cite{RCP_pq} proposed to leverage {\it oblivious transfer} to add query privacy to RCP-I and RCP-II~\cite{RCP}.  However, for similar reasons as NISAN, this scheme is vulnerable to the range estimation attack, since the initiator needs to contact multiple intermediate nodes at each step of the lookup.

Freenet~\cite{freenet} is a deployed P2P system, which allows people to upload sensitive files to the overlay and employs data duplication strategies to make them hard to block. Freenet aims to preserve the publishers' privacy, but does not provide anonymity in lookups. Vasserman et al.~\cite{mcon} create a membership concealing overlay network (MCON) for unobservable communication. They aim to make it difficult for either an insider or outsider adversary to learn the set of participating members. This is similar to previous {\it darknet} designs~\cite{waste}. However, MCON and darknets are not designed to provide anonymity.

%% file: conclusion.tex
\section{Conclusion}
\label{sec:conclusion}

In this paper, we presented Octopus, a new DHT lookup that provides strong guarantees for both anonymity and security. Octopus ensures security and anonymity via three fundamental techniques. First, Octopus constructs an anonymous path to send lookup query messages while hiding the initiator. Second, it splits the individual queries used in a lookup over multiple paths, and introduces dummy queries, to make it difficult for an adversary to learn the lookup target. Third, it uses secret security checks to identify and remove malicious nodes.  We developed an event-based simulator, and showed that malicious nodes can be quickly identified with high accuracy. In addition, via probabilistic modeling and simulation, we showed that Octopus can achieve near-optimal anonymity for both the lookup initiator and target. We also evaluated the efficiency of Octopus on Planetlab, and showed that Octopus has reasonable lookup latency and communication overhead.

%% file: appendix.tex
%\newpage 

\appendix
\label{appendix}

\subsection{Random Walk for Relay Selection}
\label{ssec:random_walk}

\begin{figure}[h]
        \centering
        \includegraphics[width=8.5cm]{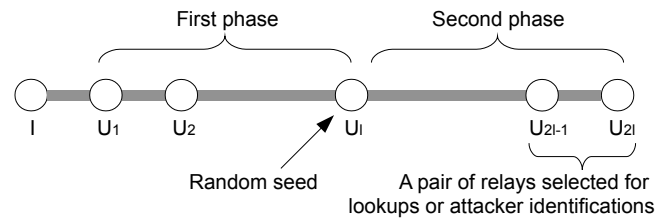}
        \caption{Two-phase random walk for selecting relays. } 
        \label{fig:random_walk}
\end{figure}

As shown in Figure \ref{fig:random_walk}, the random walk originates from the initiator $I$ and is composed of two phases, with $l$ nodes visited in each phase ($l = \Theta\log(N)$). The motivation of dividing the random walk into two phases is to mitigate the timing analysis attack, wherein malicious nodes on the same anonymous path can be associated by analyzing timings of packets in the traffic going through them.  In the first phase, $I$ picks a random finger $U_1$ out of its fingertable, and requests $U_1$ for its fingertable, from which the second hop $U_2$ is selected. Then $I$ sends an onion-encrypted query to $U_2$, using $U_1$ as the forwarding node, and selects the third hop $U_3$ at random from the fingertable returned by $U_2$. This process is recursively repeated for $l$ hops. To provide integrity check and source authentication, each replied fingertable is signed by its owner with the owner's certificate attached. 

The second phase of the random walk is conducted by $U_l$, the last node visited in the first phase. In particular, $I$ sends $U_l$ a random seed through the anonymous path established in the first phase, and the seed will guide $U_l$ how to pick nodes ``randomly''. For example, we can let $U_l$ apply hash function to the seed for $i$ times and map the hash value to $[1, m]$ ($m$ is the size of fingertables) and use the result as an index to select the $i$-th hop. The second phase is performed in the same way as the first phase, and the last two hops ($U_{2l-1}$ and $U_{2l}$) are chosen as a pair of relays to be used in lookups or attacker identifications.
To prevent malicious $U_l$ from biasing the random walk, $U_l$ is required to keep all received fingertables, signatures and certificates, and send them back to $I$ through the anonymous path of the first phase at the end of the random walk. Such information allows $I$ to verify whether $U_l$ has honestly performed the random walk. If the verification is invalid or $I$ does not receive the results by a pre-set deadline, $I$ chooses another node from $U_{l-1}$'s fingertable to restart the second phase of the random walk.

\subsection{Other Attacks, Countermeasures, and Discussion}
\label{ssec:other_attacks}

\subsubsection{Selective Denial-of-Service Attack}

A threat to anonymous communication systems (like Tor~\cite{tor}), the selective Denial-of-Service (DoS) attack~\cite{Borisov07},  can increase the chance of compromising anonymous circuits by selectively dropping packets to tear down the circuits that are infeasible to compromise. The selective DoS attack is also applicable to Octopus. For example, to create more opportunities of observing lookup initiators, malicious relays can selectively drop lookup queries or replies, when the relay directly connected to the initiator is not malicious. Nevertheless, under our framework of attacker identification, Octopus can effectively constrain the selective DoS attack by identifying malicious droppers.

We leverage the reputation-based reliability enhancement strategy for mix networks~\cite{Roger_reputation} to identify malicious dropper nodes. The idea is as follows. 
Each message is assigned a deadline by which it must be sent to the next hop (in either direction along the anonymous path). A relay $X$ first tries to send the message to the next hop $Y$ directly. If $Y$ is alive and honest, it will send a signed {\it receipt} back to $X$. If $X$ has not received a receipt from $Y$ by a specified period before the deadline, it will request a pre-defined set of {\it witnesses} (e.g., its successors and predecessors) to independently try to send the message to $Y$ and obtain a receipt. If a witness gets a valid receipt, it will forward it to $X$; otherwise, it sends $X$ a signed {\it statement} to the delivery failure. 
Before sending any query, the initiator first checks if the first pair of relays $A$, $B$ are alive ($B$ is checked by $A$).
During the lookup, if the initiator does not receive the $i$-th query reply by the pre-set deadline, it queries the successors and predecessors of the relays $C_i$, $D_i$ (through the partial anonymous path $A$ and $B$) about their aliveness, which can be inferred based on their recent stabilization activities. If both of them are alive, the initiator
reports the failure to the CA with the identities of all the relays. Then the CA will request the relays to provide either receipts or statements, and based on the provided information, the CA will be able to identify the malicious dropper node.

The (selective) DoS attacks are also possible in random walks. For example, a malicious hop of the random walk could simply drop packets to prevent the random walk from being completed, or a malicious $U_l$ could deny returning the random walk result to the initiator if the result only contains honest nodes. We can use the same strategy as above to identify such malicious nodes. 

\begin{figure}[h]
        \centering
        \includegraphics[height=3cm]{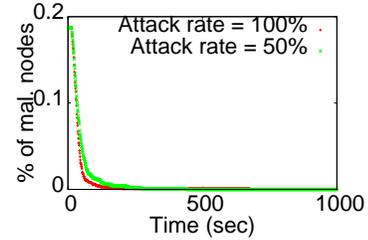}
        \caption{Selective DoS attack}
        \label{fig:dos_attack}
\end{figure}

We use the event-based simulator (described in Section~\ref{sssec:exp_setup}) to evaluate the defense mechanism for the selective DoS attack. The simulation results are shown in Figure~\ref{fig:dos_attack}. We can see that the malicious dropper nodes can be rapidly discovered by our identification mechanism.

\subsubsection{Relay Exhaustion Attack}

The relay exhaustion attack~\cite{qiyan10} is a selective-DoS-flavored attack used to compromise DHT-based anonymity systems that are lack of target anonymity protection. In such an attack, an adversary can utilize the information leak about the lookup target to predict the next relay of the circuit and launch flooding-based attack to prevent the circuit from being extended to the next relay. While Octopus also uses relays in the lookup, it is resistant to the relay exhaustion attack, since little information about the target is leaked in Octopus (as shown in Section~\ref{ssec:anonymity_evaluation}).

\subsubsection{End-to-End Timing Analysis Attack}

The end-to-end timing analysis attack is an attack that associates two malicious relays  (e.g., $A$ and $D_i$ in Figure~\ref{fig:multi_paths}) on the same anonymous path, by analyzing timings of packets in the traffic going through them. Since in Octopus there is only one message transmitted through each anonymous path in either forward or backward direction, the timing analysis attacks that require a large number of observed packets (such as packet counting~\cite{packet_counting} or packet timing correlation~\cite{watermark_oakland07, traffic_analysis_PET04}) are inapplicable to Octopus.
  
For Octopus, the only strategy to associate $A$ and $D_i$ is using the similarity of upstream and downstream latencies: in a noise-free network environment, the transmission latency from $A$ to $D_i$ should be the same as that from $D_i$ to $A$. However, this similarity is determined by communication latency jitters and can be easily destroyed by adding a short random delay at the middle relay $B$. We simulate this attack using King dataset \cite{king} to find a pair of $A$ and $D_i$ with the smallest difference between the upstream and downstream latencies. A random delay is added at $B$ from the range $[0, T_m]$, where $T_m$ is the maximum delay. We choose a typical network setting as used in related works~\cite{qiyan10, prateek08, shadowwalker, NISAN}: there are $N = 1\,000\,000$ nodes in the network, 20\% of them are malicious nodes, and concurrent lookup rate $a$ is between 0.5\% - 5\%; the jitter window for a pair of communicating peers is set as 10 ms or 10\% of the averaged transmission latency whichever is smaller (according to~\cite{n2n_delay}). 
Table \ref{tab:false_positive} presents the simulation results of this attack. When the maximum delay $T_m$ is 100 ms and $a = 5\%$, the error rate is as high as 99.91\%; in this case the information leak is only $(1-99.91\%) \cdot \log_2 (N\cdot 0.8 + N\cdot \alpha \cdot 0.2) = 0.018$ bit. This means the adversary can hardly learn extra information by launching the timing analysis attack.

\begin{table} [h] 
		\centering        
		\caption{Error rate of end-to-end timing analysis attack. $a$ is concurrent lookup rate.}
        \label{tab:false_positive}

        \begin{tabular}{|c|c|c|c|}	\hline                
		Max. delay   & $a=0.5\%$        & $a=1\%$         	& $a=5\%$	\\	\hline                
		100 ms 		& 99.35\%        	& 99.50\%        	& 99.91\%            \\	\hline
		200 ms 		& 99.60\%        	& 99.82\%        	& 99.95\%            \\	\hline		
        \end{tabular}

\end{table}

%% file: main.bbl
\begin{thebibliography}{10}
\providecommand{\url}[1]{#1}
\csname url@samestyle\endcsname
\providecommand{\newblock}{\relax}
\providecommand{\bibinfo}[2]{#2}
\providecommand{\BIBentrySTDinterwordspacing}{\spaceskip=0pt\relax}
\providecommand{\BIBentryALTinterwordstretchfactor}{4}
\providecommand{\BIBentryALTinterwordspacing}{\spaceskip=\fontdimen2\font plus
\BIBentryALTinterwordstretchfactor\fontdimen3\font minus
  \fontdimen4\font\relax}
\providecommand{\BIBforeignlanguage}[2]{{%
\expandafter\ifx\csname l@#1\endcsname\relax
\typeout{** WARNING: IEEEtran.bst: No hyphenation pattern has been}%
\typeout{** loaded for the language `#1'. Using the pattern for}%
\typeout{** the default language instead.}%
\else
\language=\csname l@#1\endcsname
\fi
#2}}
\providecommand{\BIBdecl}{\relax}
\BIBdecl

\bibitem{chord}
I.~Stoica, R.~Morris, D.~Liben-Nowell, D.~R. Karger, M.~F. Kaashoek, F.~Dabek,
  and H.~Balakrishnan, ``Chord: {A} scalable peer-to-peer lookup protocol for
  internet applications,'' \emph{IEEE/ACM Trans. Netw.}, vol.~11, no.~1, pp.
  17--32, 2003.

\bibitem{Kad}
P.~Maymounkov and D.~Mazieres, ``Kademlia: A peer-to-peer information system
  based on the xor metric,'' \emph{IPTPS}, 2001.

\bibitem{freedman+:nsdi04}
M.~J. Freedman, E.~Freudenthal, and D.~Mazi\`{e}res, ``Democratizing content
  publication with coral,'' in \emph{Proc. of NSDI'04}, 2004.

\bibitem{rowstron-druschel:sosp01}
A.~Rowstron and P.~Druschel, ``Storage management and caching in past, a
  large-scale, persistent peer-to-peer storage utility,'' in \emph{Proc.
  SOSP'01}, 2001.

\bibitem{AP3}
A.~Mislove, G.~Oberoi, A.~Post, C.~Reis, P.~Druschel, and D.~S. Wallach, ``Ap3:
  Cooperative, decentrialized anonymous communication,'' \emph{ACM SIGOPS
  European Workshop}, 2004.

\bibitem{salsa}
A.~Nambiar and M.~Wright, ``Salsa: A structured approach to large-scale
  anonymity,'' \emph{ACM CCS}, 2006.

\bibitem{NISAN}
A.~Panchenko, S.~Richter, and A.~Rache, ``Nisan: Network information service
  for anonymization networks,'' \emph{ACM CCS}, November 2009.

\bibitem{torsk}
J.~McLachlan, A.~Tran, N.~Hopper, and Y.~Kim, ``Scalable onion routing with
  torsk,'' \emph{ACM CCS}, November 2009.

\bibitem{shadowwalker}
P.~Mittal and N.~Borisov, ``Shadowwalker: Peer-to-peer anonymous communication
  using redundant structured topologies,'' \emph{ACM CCS}, November 2009.

\bibitem{shakimov+:wosn09}
A.~Shakimov, A.~Varshavsky, L.~P. Cox, and R.~C\'{a}ceres, ``Privacy, cost, and
  availability tradeoffs in decentralized osns,'' ser. WOSN '09, 2009, pp.
  13--18.

\bibitem{social_cutillo}
L.~A. Cutillo, R.~Molva, and T.~Strufe, ``Privacy preserving social networking
  through decentralization,'' \emph{Proceedings of the Sixth international
  conference on Wireless On-Demand Network Systems and Services}, 2009.

\bibitem{wallach:ssts03}
D.~Wallach, ``A survey of peer-to-peer security issues,'' in \emph{Software
  Security — Theories and Systems}, ser. LNCS, 2003, vol. 2609, pp. 253--258.

\bibitem{prateek08}
P.~Mittal and N.~Borisov, ``Information leaks in structured peer-to-peer
  anonymous communication systems,'' \emph{ACM CCS}, 2008.

\bibitem{qiyan10}
Q.~Wang, P.~Mittal, and N.~Borisov, ``In search of an anonymous and secure
  lookup: Attacks on structured peer-to-peer anonymous communication systems,''
  \emph{ACM CCS}, 2010.

\bibitem{castro}
M.~Castro, P.~Druschel, A.~Ganesh, A.~Rowstron, and D.~S. Wallach, ``Secure
  routing for structured peer-to-peer overlay networks,'' in \emph{OSDI},
  December 2002.

\bibitem{Halo}
A.~Kapadia and N.~Triandopoulos, ``Halo: High-assurance locate for distributed
  hash tables,'' in \emph{NDSS}, February 2008.

\bibitem{RCP}
M.Young, A.Kate, I.Goldberg, and M.Karsten, ``Practical robust communication in
  dhts tolerating a byzantine adversary,'' \emph{Proc. ICDCS'10}, pp. 263--272,
  2010.

\bibitem{RCP_pq}
M.~Backes, I.Goldberg, A.~Kate, and T.~Toft, ``Adding query privacy to robust
  dhts,'' \emph{Tech. rep., arXiv:1107.1072v1 [cs.CR]}, July 2011.

\bibitem{sybil}
J.~Douceur, ``The sybil attack,'' in \emph{Peer-to-Peer Systems}, 2002, vol.
  2429, pp. 251--260.

\bibitem{nikita_sybil}
N.~Borisov, ``Computational puzzles as sybil defenses,'' in \emph{Peer-to-Peer
  Computing}, 2006.

\bibitem{Sybil_Geroge_09}
G.~Danezis and P.~Mittal, ``Sybilinfer: Detecting sybil nodes using social
  networks,'' in \emph{NDSS}, 2009.

\bibitem{Sybillimit}
H.~Yu, P.~B. Gibbons, M.~Kaminsky, and F.~Xiao, ``Sybillimit: A near-optimal
  social network defense against sybil attacks,'' in \emph{IEEE Symposium on
  Security and Privacy (Oakland)}, 2008.

\bibitem{anon_terminology}
A.~Pfitzmann and M.~Hansen, ``A terminology for talking about privacy by data
  minimization: Anonymity, unlinkability, undetectability, unobservability,
  pseudonymity, and identity management,'' Aug. 2010, v0.34.

\bibitem{Pastry}
A.~Rowstron and P.~Druschel, ``Pastry: Scalable, decentralized object location
  and routing for large-scale peer-to-peer systems,'' \emph{IFIP/ACM
  International Conference on Distributed Systems Platforms (Middleware)}, pp.
  329--350.

\bibitem{Myrmic}
P.~Wang, I.~Osipkov, N.~Hopper, and Y.~Kim, ``Myrmic: Secure and robust dht
  routing,'' \emph{Tech. rep., Digital Technology Center, University of
  Minnesota at Twin Cities}, 2007.

\bibitem{revocation-tree}
J.~L. Muñoz, J.~Forne, O.~Esparza, and M.~Soriano, ``Certificate revocation
  system implementation based on the merkle hash tree,'' \emph{Inte. Journ. of
  Inf. Sec.}, vol.~2, no.~2, 2003.

\bibitem{revocation-p2p}
M.~C. Morogan and S.~Muftic, ``Certificate revocation system based on
  peer-to-peer crl distribution,'' in \emph{Proc. of the DMS'03 Conference},
  2003.

\bibitem{PKI-p2p}
C.~Y. Liau, S.~Bressan, K.-L. Tan, C.~Yee, L.~Stéphane, and B.~K. lee Tan,
  ``Efficient certificate revocation: A p2p approach,'' in \emph{HICSS'05},
  2005.

\bibitem{onion-routing}
P.~Syverson, G.~Tsudik, M.~Reed, and C.~Landwehr, ``Towards an analysis of
  onion routing security,'' \emph{International Workshop on Design Issues in
  Anonymity and Unobservaility, vol. 2009, LNCS, Springer}, July 2000.

\bibitem{king}
F.~Dabek, J.~Li, E.~Sit, J.~Robertson, M.~F. Kaashoek, and R.~Morris,
  ``Designing a dht for low latency and high throughput,'' \emph{NSDI'04},
  2004.

\bibitem{octopus_tr}
Q.~Wang and N.~Borisov, ``Octopus: A secure and anonymous dht lookup,'' uiuc-cs
  technical report, 2011.
  \url{http://hatswitch.org/~qwang26/papers/octopus-tr.pdf}.

\bibitem{Cyclone}
M.~S. Artigas, P.~G. Lopez, J.~P. Ahullo, and A.~F.~G. Skarmeta, ``Cyclone: A
  novel design schema for hierarchical dhts,'' \emph{In P2P ’05}, pp. 49--56,
  2005.

\bibitem{eclipse_attack}
M.~Schuchard, A.~Dean, V.~Heorhiadi, N.~Hopper, and Y.~Kim, ``Balancing the
  shadows,'' \emph{ACM WPES}, 2010.

\bibitem{freenet}
I.~Clarke, T.~W. Hong, S.~G. Miller, O.~Sandberg, and B.~Wiley, ``{Protecting
  Free Expression Online with \{Freenet\}},'' \emph{IEEE Internet Computing},
  vol.~6, no.~1, pp. 40--49, 2002.

\bibitem{mcon}
E.~Y. Vasserman, R.~Jansen, J.~Tyra, N.~Hopper, and Y.~Kim,
  ``Membership-concealing overlay networks,'' in \emph{ACM CCS}, 2009.

\bibitem{waste}
WASTE. \url{http://waste.sourceforge.net/}.

\bibitem{packet_counting}
A.~Serjantov and P.~Sewell, ``Passive attack analysis for connection-based
  anonymity systems,'' in \emph{ESORICS'03}, October 2003.

\bibitem{watermark_oakland07}
X.~Wang, S.~Chen, and S.~Jajodia, ``Network flow watermarking attack on
  low-latency anonymous communication systems,'' in \emph{IEEE Security and
  Privacy}, May 2007.

\bibitem{traffic_analysis_PET04}
G.~Danezis, ``The traffic analysis of continuous-time mixes,'' in \emph{Privacy
  Enhancing Technologies workshop}, May 2004.

\bibitem{n2n_delay}
A.~Acharya and J.~Saltz, ``A study of internet round-trip delay,''
  \emph{Technical Report (CS-TR 3738), University of Maryland}, 1996.

\bibitem{tor}
R.~Dingledine, N.~Mathewson, and P.~Syverson, ``Tor: The second-generation
  onion router,'' in \emph{USENIX Security Symposium}, August 2004.

\bibitem{Borisov07}
N.~Borisov, G.~Danezis, P.~Mittal, and P.~Tabriz, ``Denial of service or denial
  of security?'' \emph{ACM CCS}, 2007.

\bibitem{Roger_reputation}
R.~Dingledine, M.~J. Freedman, D.~Hopwood, and D.~Molnar, ``A reputation system
  to increase mix-net reliability,'' in \emph{the 4th International Workshop on
  Information Hiding (IHW)}, 2001.

\end{thebibliography}
